\documentclass[prx,twocolumn,showpacs,preprintnumbers]{revtex4-1}
\usepackage{graphicx}
\usepackage{amsfonts}
\usepackage{amssymb}
\usepackage{amsmath}
\usepackage{bm}
\usepackage{mathtools}
\usepackage{bbold}
\usepackage{color}
\usepackage{physics}
\usepackage{nicefrac}
\usepackage{pgfplots}

\newcommand{\be}{\begin{equation}}
\newcommand{\ee}{\end{equation}}
\newcommand{\bea}{\begin{eqnarray}}
\newcommand{\eea}{\end{eqnarray}}
\newcommand{\Eq}[1]{Eq.\,(\ref{#1})}
\newcommand{\Eqs}[2]{Eqs.\,(\ref{#1}) and (\ref{#2})}
\newcommand{\Fig}[1]{Fig.\,\ref{#1}}
\newcommand{\Figs}[2]{Figs.\,\ref{#1} and \ref{#2}}
\newcommand{\Sec}[1]{Sec.\,\ref{#1}}
\newcommand{\App}[1]{Appendix\,\ref{#1}}
\newcommand{\Vol}{\mathbb{V}}%

\begin{document}

\title{Phonon-induced dephasing in quantum dot-cavity QED: Limitations of the polaron master equation}

\author{A. Morreau}
\email[Electronic address: ]{morreauai@cardiff.ac.uk}
\author{C. Joshi}
\author{E. A. Muljarov}
\email[Electronic address: ]{egor.muljarov@astro.cf.ac.uk}
\affiliation{%
School of Physics and Astronomy, Cardiff University, Cardiff CF24 3AA, United Kingdom}
\date{\today}

\begin{abstract}
A semiconductor quantum dot (QD) embedded within an optical microcavity is a system of fundamental importance within quantum information processing. The optimization of quantum coherence is crucial in such applications, requiring an in-depth understanding of the relevant decoherence mechanisms. We provide herein a critical review of prevalent theoretical treatments of the QD-cavity system coupled to longitudinal acoustic phonons, comparing predictions against a recently obtained exact solution. Within this review we consider a range of temperatures and exciton-cavity coupling strengths. Predictions of the polaron Nakajima-Zwanzig (NZ) and time-convolutionless (TCL) master equations, as well as a variation of the former adapted for adiabatic continuous wave excitation (CWE), are compared against an asymptotically exact solution based upon Trotter's decomposition (TD) theorem. The NZ and TCL implementations, which apply a polaron transformation to the Hamiltonian and subsequently treat the exciton-cavity coupling to second order, do not offer a significant improvement accuracy relative to the polaron transformation alone. The CWE adaptation provides a marked improvement, capturing the broadband features of the absorption spectrum (not present in NZ and TCL implementations). We attribute this difference to the effect of the Markov approximation, and particularly its unsuitability in pulsed excitation regime. Even the CWE adaptation, however, breaks down in the regime of high temperature ($50K$) and strong exciton-cavity coupling ($g \gtrsim 0.2$ meV). The TD solution is of comparable computational complexity to the above-mentioned master equation approaches, yet remains accurate at higher temperatures and across a broad range of exciton-cavity coupling strengths (at least up to $g=1.5$ meV).
\end{abstract}

\maketitle

\section{Introduction}
The field of cavity quantum electrodynamics (cavity QED) has been traditionally associated with the interaction between a confined light field and an atomic system~\cite{JaynesIEEE63}. Following the development of semiconductor and solid-state ``artificial atoms''~\cite{Chang1974,Dingle1974}, however, a new branch of research within this field has emerged. Solid-state cavity QED devices offer significant advantages over their atomic counterparts. Notably solid state systems provide precise tunability and access to stronger coupling regimes, thereby opening the possibility of previously unobserved physical phenomena.

We focus, in this work, on the {\it strong coupling regime}, in which there is a partly reversible exchange of energy between the exciton and cavity modes. In this regime, the absorption and coherent re-emission of a photon  occurs many times before the photon leaks from the cavity or is spontaneously emitted, giving rise to {\em polariton} formation and characteristic vacuum Rabi splitting~\cite{ThompsonPRL92,ReitzensteinJPhysD10,OtaAPL18}.

The strong coupling regime of the idealized QD-cavity system introduced above is well-described by the exactly solvable Jaynes-Cummings (JC) model~\cite{JaynesIEEE63,VallePRB09,KasprzakNatMat10}. This model accounts for the exciton-cavity interaction but neglects the ever-present coupling of the QD exciton to the environment, which includes, first of all, the interaction with longitudinal acoustic (LA) phonons. There is significant experimental and theoretical evidence~\cite{Wilson-RaePRB02,McCutcheonNJP10,Ota09,KaerPRL10,HohenesterPRB10,RoyPRL11,GlasslPRB12,NazirJPCM16,NahriJPCM16,HorneckerPRB17,HohenesterPRB09,CalicPRL11,ValentePRB14,PortalupiNL15,MullerPRX15} to suggest that the phononic environment plays a crucial role in the optical decoherence of the QD-cavity system and must therefore be taken into account in any realistic model of such a system.

A number of approaches to the dissipative QD-cavity problem have been suggested in the literature, many of which are based upon the quantum master equation~\cite{Breuer06}. Whilst several variations of the master equation approach exist, all rely upon the perturbative treatment of at least one interaction term within the full dissipative QD-cavity system.


The {\it weak coupling master equation}, for example, treats the exciton-phonon interactions perturbatively to second order~\cite{RoyPRB12}. It has been applied in the Markovian~\cite{RamsayPRL10,NazirJPCM16} and non-Markovian~\cite{MachnikowskiPRB04,MogilevtsevPRL08} regimes, but, as the name suggests, is suitable only in the regime of weak exciton-phonon interactions. Moreover, it is known to break down at elevated temperatures ($\gtrsim 30$ K)~\cite{McCutcheonNJP10,NazirJPCM16}, where multi-phonon effects become significant.

For stronger exciton-phonon interactions, the {\it polaron master equation} is more appropriate. Here, the exciton-phonon coupling is assumed to be the dominant interaction, modifying the exciton mode to a {\it polaron} (phonon-dressed exciton) state. Formally, this modification is made through a polaron transformation of the system Hamiltonian, with the polaron-cavity interaction treated perturbatively to the second order Born approximation~\cite{Wilson-RaePRB02,KaerPRL10,Ota09,McCutcheonNJP10} or beyond~\cite{NazirJPCM16, HorneckerPRB17}. Unlike the weak coupling master equation, the polaron master equation accounts for multi-phonon processes~\cite{McCutcheonNJP10} and predicts a phonon-induced renormalization of the exciton-cavity coupling strength~\cite{McCutcheonNJP10}. However, it is known to fail when the exchange of energy between the QD and exciton modes occurs on a timescale shorter than, or comparable to, the time required for the phonon bath to form (or disperse) a cloud surrounding the newly formed (or annihilated) exciton~\cite{McCutcheonNJP10}.

Bridging the disparate parameter regimes of the weak coupling and polaron master equations, McCutcheon {\it et al.}~\cite{McCutcheonPRB11,NazirJPCM16} have proposed the {\it variational master equation}. This approach, inspired by variational polaron theory~\cite{SilbeyJCP84,Harris1985,Silbey1989}, employs a unitary transformation of the full system Hamiltonian similar to the polaron transformation. The key distinction, however, is that the magnitude of the phonon displacement operator is dictated by the state of the excitonic system and found through free energy minimization. Whilst providing an improved range of validity in comparison to the weak coupling and polaron master equations alone, the variational master equation is nonetheless, at its core, a perturbative technique.

Moreover, as is common with all master equation formalisms, the complex dynamical evolution of the many-body environment is not calculated explicitly. Instead, the exciton-cavity subsystem (with displaced phonon operators for the case of the polaron and variational master equations) is separated from the environment and all information relating to the evolution of the latter is lost. It has been shown~\cite{Iles-Smith2014} that proper treatment of the phonon environment is crucial in order to correctly describe both the system transient and equilibrium behavior of the exciton-cavity subsystem. Whilst certain measures have been proposed to mitigate this limitation, notably reaction coordinate mapping~\cite{Iles-Smith2014}, such measures complicate the approach and hence obviate the master equation's core advantages of relative simplicity and intelligibility.

In addition to the above-described master equation formalisms, a number of non-perturbative approaches have been proposed in the literature. Feynman's path integral formulation forms the basis of many such methods, with summation over all possible paths implemented numerically~\cite{MakarovCPL94,MakriJCP95a,MakriJCP95b,NahriJPCM16,GlasslPRB12,VagovPRL07,VagovPRB11,Barth2016,Strathearn2017}. These methods provide numerical convergence to an exact solution but are computationally expensive and offer little by way of physical insight.

The Trotter decomposition (TD) method with linked cluster expansion~\cite{MorreauPRB19} builds upon the path integral formalism. In this technique, the full dissipative qubit-cavity system is separated into two exactly solvable subsystems, described by the JC and independent boson (IB) models respectively. Notably, solution of the latter via the linked cluster
expansion~\cite{Mahan00} enables the summation over all possible paths to be implemented through matrix multiplication.


In this work we examine the validity of three variations of the polaron master equation, implementing each method across a range of key parameters including temperature and qubit-cavity coupling strength. We employ the TD to provide exact solutions against which to measure the accuracy of the master equation techniques. As we will show, the TD shares many parameters with the polaron master equation techniques, and hence the TD is a natural method to draw upon in order to obtain exact results. To ease the comparison, we concentrate on the linear optical polarization. This allowed us to bring all three models based on the polaron mater equation to a fully analytic form, enabling their explicit analysis and a direct comparison with each other and with the exact solution.

The paper is organized as follows: in \Sec{sec:Hamiltonian} we introduce the Hamiltonian of the dissipative QD-cavity system, alongside the Lindblad master equation and characteristic timescales. An overview of the excitation regimes and associated expressions for linear optical absorption is provided in \Sec{sec:overview}. This is followed, in \Sec{sec:methods}, with a detailed outline of the three polaron master equations and TD. Finally, the absorption spectra calculated according to the various methods are compared in \Sec{sec:comparison}, with conclusions drawn in \Sec{sec:conclusion}.

\section{System Hamiltonian and characteristic timescales}
\label{sec:Hamiltonian} 
A semiconductor quantum dot (QD) confined within an optical microcavity constitutes an excellent environment in which to study the nature of light-matter interaction at a quantum-mechanical level. The discretization of the electronic band structure ensures that the promotion of an electron from the valence to the conduction band (and associated generation of a hole) occurs at a distinct frequency known as the exciton transition frequency $\omega_X$.

One may, in general, restrict the electron-hole dynamics to a two-level system (TLS) consisting of states $\ket{0}$ and $\ket{X}$, representing the fully unexcited system and the exciton ground state respectively. The interaction of this two-level fermionic system with the single optical mode of the confining microcavity $\ket{C}$ (of frequency $\omega_C$) provides significant insight into the underlying physics of the light-matter interaction. Moreover, such a system is potentially suitable for solid state quantum information processing and therefore is of practical and technological importance.

Treating the phononic environment according to the standard harmonic oscillator model, we arrive at the following full system Hamiltonian $\mathcal{H}$ (in the units of $\hbar=1$)
\begin{equation}
\mathcal{H} = \omega_X d^\dagger d + \omega_C a^\dagger a + g(a^\dagger d + d^\dagger a) + \mathcal{H}_{\rm ph} + d^\dagger d V, \label{eq:Hlab}
\end{equation}
where $d^{\dagger}$ ($a^{\dagger}$) is the exciton (cavity photon) creation operator, $g$ is the exciton-cavity coupling strength, $\mathcal{H}_{\rm ph}$ is the free phonon bath Hamiltonian, and $V$ is the exciton-phonon interaction. The latter two entities may be expressed in terms of the creation operator $b_q^{\dagger}$, energy $\omega_q$ and the matrix element $\lambda_q$ of the coupling of the $q$-th phonon mode to the QD exciton,
\begin{equation}
\mathcal{H}_{\rm ph} = \sum_{q}\omega_q b_q^\dagger b_q\,,\ \ \ \
V = \sum_q \lambda_q (b_q + b_{-q}^\dagger)\,.
\label{eq:HIBcomponents}
\end{equation}

Whilst not present within this Hamiltonian, we allow for the Markovian dephasing processes of radiative decay $\gamma_C$ and long-time ZPL dephasing $\gamma_X$. These processes are taken into account through the {\it Lindblad dissipator} $\mathcal{D}$ within the master equation formalism,
\begin{align}
i\dot{\rho}&=[{\cal H},\rho] + \mathcal{D},\label{eq:Lind_master}\\
\mathcal{D}&=i\gamma_X\left(2d\rho d^{\dagger} - d^{\dagger}d\rho - \rho d^{\dagger} d\right)\nonumber\\
&\hspace{0.5cm}+ i\gamma_C\left(2a\rho a^{\dagger} - a^{\dagger}a\rho - \rho a^{\dagger} a\right)\,,\label{eq:Lind_diss}
\end{align}
where $\rho(t)$ is the density matrix of the full exciton-cavity-phonon system and $\dot{\rho}$ is its time derivative.

It should be noted that the system Hamiltonian $\mathcal{H}$ does not vary with time. Accordingly, the energy eigenstates (defined by the relation $\mathcal{H}\ket{n} = E_n\ket{n}$) are stationary states. We may, however, apply an external perturbation $\mathcal{V}(t)$ to induce inter-state transitions. The rate of such transitions as a function of energy is directly observable through the absorption spectrum.

There are two key timescales within the full exciton-cavity-phonon system. The first, $\tau_{\rm JC}$, is associated with the frequency of Rabi oscillations between the exciton and cavity modes. This is approximately inversely proportional to the exciton-cavity coupling strength $g$,
\begin{equation}
\tau_{\rm JC} \sim \frac{\pi}{g}\,.\label{eq:tauJC}
\end{equation}
The second timescale describes the phonon memory time $\tau_{\rm IB}$. This timescale is independent of the exciton-cavity coupling strength $g$ but varies according to other system parameters. For temperatures $\gtrsim 5$ K, the polaron timescale $\tau_{\rm IB}$ may be approximated as~\cite{MorreauPRB19}
\begin{equation}
\tau_{\rm IB} \approx \sqrt{2}\pi l/v_s\,,
\label{eq:tauIB}
\end{equation}
where $l$ is the exciton confinement radius and $v_s$ is the speed of sound in the QD material.

\section{Overview: polarization and absorption }
\label{sec:overview}
In this paper we focus on the simple and intuitively clear quantities of linear optical polarization and absorption. With this in mind, we define a density matrix of the exciton-cavity subsystem $\rho_S(t) = \tr_B\{\rho(t)\}$, where the trace is taken over all phonon states. Our restriction to the linear regime allows reduction of the QD-cavity basis to the following three states: the absolute ground state $\ket{0}$, the exciton mode $\ket{X}$, and the cavity mode $\ket{C}$. In this basis, $d^{\dagger} = \ket{X}\bra{0}$, $a^{\dagger} = \ket{C}\bra{0}$ and the QD-cavity subsystem density matrix $\rho_{S}(t)$ has the form
\begin{equation}
\rho_{S}(t) = \sum_{m,n=0,X,C} \rho_{mn} (t)\ket{m} \bra{n}\,.\label{eq:rhos}
\end{equation}
The phononic contribution to the full density matrix $\rho(t)$ is treated with differing rigor within the approaches described below.

In terms of external excitation $\mathcal{V}(t)$, we consider two contrasting regimes: pulsed excitation and weak adiabatic continuous wave excitation (CWE). In each case, we allow for excitation in the cavity mode $\ket{C}$ or in the exciton mode $\ket{X}$.

\subsection{Continuous wave excitation}
\label{sec:CWoverview}
A continuous wave excitation, switched on at time $t_0$ has the form
\begin{equation}
\mathcal{V}_{\rm CW}(t) =  e^{\epsilon t} \, \Theta(t-t_0) \, \sum_e \Omega_e \left(c_e^{\dagger} \, e^{-i\omega t} + c_e \, e^{i\omega t} \right)\,, \label{eq:VCW}
\end{equation}
where $\Omega_e$ is a constant that parameterizes the strength of the excitation and $c_e^{\dagger}$ is the creation operator associated with the feeding channel: $c_e^{\dagger}=a^{\dagger}$,  $c_e^{\dagger}=d^{\dagger}$ or both together, for excitation in the cavity, exciton, or both modes, respectively.

Once switched on, the excitation $\mathcal{V}(t)$ perturbs the system such that a general state $\ket{\psi(t)}$ evolves as $\ket{\psi(t)} = \sum_n a_n(t) \ket{n}$. The coefficients $a_n(t)$ are given by standard time-dependent perturbation theory~\cite{Sakurai14} to first order as
\begin{equation}
a_n(t) = \delta_{in} - i \int_{t_0}^t dt' \, \bra{n} \tilde{\mathcal{V}}_{\rm CW}(t') \ket{i}\,, \label{eq:cf_full}
\end{equation}
where $\ket{i}$ is the initial state of the system prior to excitation and the tilde notation indicates the interaction representation, defined as
\begin{equation}
\tilde{\mathcal{V}}_{\rm CW}(t) = e^{i \mathcal{H} t} \, \mathcal{V}_{\rm CW}(t) \, e^{-i\mathcal{H}t}\,.
\end{equation}

With the physical observable of absorption in mind, we wish to find the probability $\mathcal{P}_f(t) = |a_f(t)|^2$ of the system being in a specific final energy eigenstate $\ket{f}$. The absorption $A(\omega)$ then follows from the transition rate, i.e. the transition probability per unit time. To zeroth order, we see from \Eq{eq:cf_full} that $a_i = 1$. We may therefore neglect all transitions other than those that originate from the initial state $\ket{i}$,
\begin{equation}
A(\omega) = \frac{d}{dt} \sum_{f \neq i} |a_f(t)|^2 \,.\label{eq:abs_sum_Pm}
\end{equation}
 From the normalization condition $\bra{\psi(t)}\ket{\psi(t)} = \sum_n |a_n(t)|^2 = 1$, we may recast \Eq{eq:abs_sum_Pm} as
\begin{equation}
A(\omega) = - \frac{d |a_i(t)|^2}{dt} = - \frac{d}{dt} \bra{i}\rho(t)\ket{i}\,, \label{eq:abs_cl}
\end{equation}
where $\rho(t) = \ket{\psi(t)}\bra{\psi(t)}$ is the full system density matrix.
Clearly, $A(\omega)$ also depends on time $t$. However, in the limit $t\to\infty$ the pumped system reaches its equilibrium, and $A(\omega)$ becomes the time-independent steady-state absorption.

We now look to apply \Eq{eq:abs_cl} to the system in hand, namely a qubit-cavity system coupled to a phonon bath. We assume that the phonon bath is initially in a thermal state and that $\omega_X, \, \omega_C \gg k_B T$ [hence the exciton-cavity subsystem is in its absolute ground state $\ket{0}$]. The initial system is therefore described by density matrix $\rho(-\infty) = \ket{i}\bra{i}$ of the form
\begin{align}
\rho(-\infty) &= \ket{0} \bra{0} \otimes \rho_{B}\,,\label{eq:rho_neg_inf}\\
\rho_B &=\frac{e^{-\beta \mathcal{H}_{\rm ph}}}{\tr_B \left\{e^{-\beta \mathcal{H}_{\rm ph}}\right\}}\,.
\end{align}
Here, $\otimes$ denotes the direct product, $\beta=(k_B T)^{-1}$, and the trace is taken over all possible phonon states. Assuming that the density matrix $\rho(t)$ is factorizable at all times, $\rho(t) = \rho_{S}(t) \otimes \rho_{B}(t)$, \Eq{eq:abs_cl} may be simplified to
\begin{equation}
A(\omega) = - \frac{d \rho_{00}(t)}{dt}\,,\label{eq:abs_CWE2}
\end{equation}
where $\rho_{00}(t) = \bra{0}\rho_{S}(t)\ket{0}$ with $\rho_S(t)= \tr_B\{\rho(t)\}$. 

The factorization of the density matrix used above is an approximation which is widely exploited in master equation approaches. We note however that this approximation is valid (i.e. yields asymptotically exact results) only in the so called adiabatic limit of infinitesimal excitation coupling parameters $\Omega_e\to 0$.

\subsection{Pulsed excitation}
A pulsed excitation applied at $t=t_0$ has the general form
\begin{equation}
\mathcal{V}_{\delta}(t) = \delta(t-t_0)\, \Theta_e (c_e^{\dagger} +c_e)\,, \label{eq:pulsed_exc}
\end{equation}
where $\Theta_e$ is the pulse area of the excitation of the mode $e$, and $c_e^{\dagger}$ is the same as in \Eq{eq:VCW}. Without loss of generality, we may assume that the excitation is applied at time $t_0 = 0$. The density matrix immediately after this time has the form
\begin{equation}
\rho(0_+) = e^{-i\Theta_e (c_e^{\dagger} +c_e)} \rho(-\infty) e^{i\Theta_e (c_e^{\dagger} +c_e)}, 
\label{eq:rho0+}
\end{equation}
where $\rho(-\infty)$ is the density matrix of system prior to  the excitation being applied, given in \Eq{eq:rho_neg_inf}. Note that, in contrast to the CWE regime, we account for the action of the pulsed excitation $\mathcal{V}_{\delta}(t)$ through the instantaneous evolution of the density matrix described by \Eq{eq:rho0+}; subsequent evolution of the system depends only on the time-independent Hamiltonian $\mathcal{H}$ defined in \Eq{eq:Hlab}.

The optical polarization $P(t)$ is determined by the subsequent temporal evolution of the system, encapsulated within the full density matrix $\rho(t)$,
\begin{equation}
P(t) = \tr \left\{\rho(t) c_o \right\}, \label{eq:genericP}
\end{equation}
where $\tr$ indicates the trace over all states, and the annihilation operator $c_o$ represents the observation mode; $c_o=d$ ($a$) for observation in the exciton (cavity) mode. The linear polarization $P_L(t)$ is the part of the full polarization \Eq{eq:genericP} which is linear in the pulse area $\Theta_e$.

The absorption $A(\omega)$ in the CWE regime, given by \Eq{eq:abs_cl} is straightforwardly related to the linear polarization $P_L(t)$ of the pulsed regime through the inverse Fourier transform:
\begin{equation}
A(\omega) = \Re \int_{-\infty}^{\infty} d t \, P_L(t) \, e^{i\omega t}\,,
\label{eq:Absorp_pulsed}
\end{equation}
see  \App{sec:pulsed_abs} for details. 


\section{Substantive method comparison}
\label{sec:methods}
We consider, within this comparative study, four different methods for determining the linear optical absorption of an exciton-cavity system coupled to a phononic environment. In one such approach [outlined in detail in \Sec{sec:WR}], the system is excited through a weak adiabatic CWE; in the remaining three approaches [Secs. \ref{sec:NZ}, \ref{sec:TCL} and \ref{sec:Trotter}], excitation is achieved through a spectrally broad delta pulse. As discussed in \Sec{sec:overview} above, these different excitation regimes provide identical absorption spectra.

\begin{figure}
\includegraphics{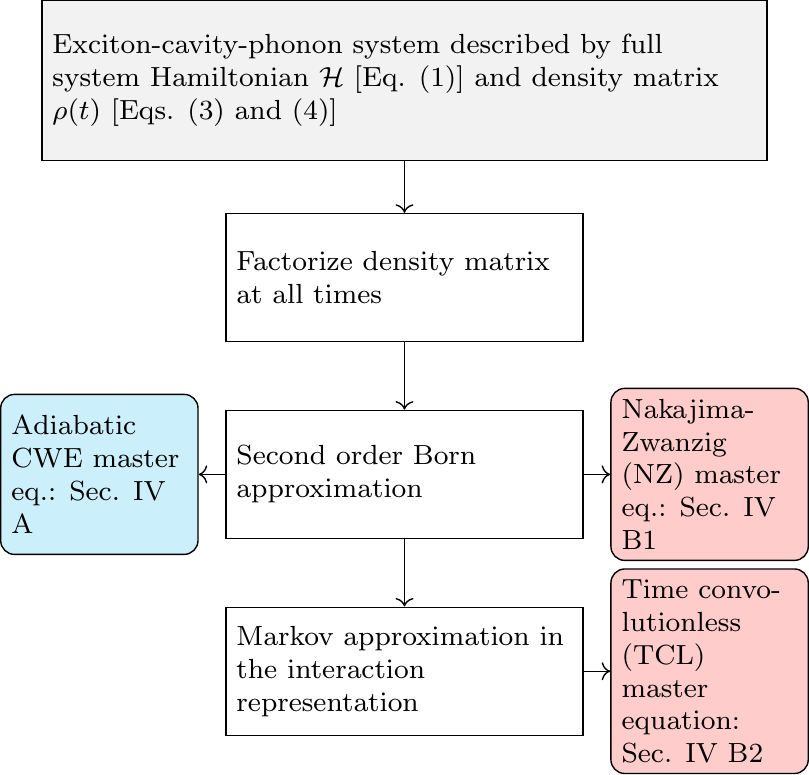}
%
%
\vspace{0.5cm}
\caption{Diagrammatic representation of the different levels of approximation associated with the three polaron master equation methods under consideration in the present work. White boxes indicate approximations or assumptions; red (blue) boxes indicate pulsed excitation (CWE) regimes. The Trotter decomposition method with linked cluster expansion [\Sec{sec:Trotter}], when applied in relation to a pulsed excitation, requires factorization of the density matrix prior to ($t<0$) and immediately after ($t=0_+$) the pulsed excitation is applied, but no further approximations or assumptions.}
\label{fig:master_eq_approximations}
\end{figure}

An overview of the relationship between the various approaches, and the associated conditions for validity, is provided in \Fig{fig:master_eq_approximations}.

We initially consider absorption in the adiabatic CWE regime, adopting a polaron master equation approach with second order Born approximation. This method was originally proposed by Wilson-Rae and Imamo{\u{g}}lu~\cite{Wilson-RaePRB02} and is discussed in detail in \Sec{sec:WR}.

We then proceed to discuss the pulsed excitation regime. The Nakajima-Zwanzig (NZ) master equation~\cite{Breuer06,RoyPRB12}, outlined in \Sec{sec:NZ} relies upon the same approximations as the above-described approach by Wilson-Rae and Imamo{\u{g}}lu. However, the validity of these approximations, and hence the accuracy of the calculated absorption spectra, may differ between the two excitation regimes.

The time-convolutionless (TCL) master equation, discussed further in \Sec{sec:TCL}, initially follows the same procedure as the NZ master equation. However, unlike the NZ equation, the TCL equation additionally relies upon the Markov approximation, thereby removing the memory from the system giving a time-local equation. Evolution of the density matrix to be solved in the time domain, with polarization and subsequently absorption determined according to Eqs. (\ref{eq:genericP}) and (\ref{eq:Absorp_pulsed}) respectively.

The fourth and final method under consideration is the Trotter decomposition method with cumulant expansion~\cite{MorreauPRB19}, which is described in detail in \Sec{sec:Trotter}. Fundamental to this approach is the separation of the full Hamiltonian $\mathcal{H}$ into the sum of two exactly solvable component parts, $\mathcal{H}_{\rm JC} + \mathcal{H}_{\rm IB}$, associated with the JC and IB models respectively. Trotter's decomposition theorem enables separation of the temporal evolution of the density matrix into short time slices, each wholly determined by either the JC or IB Hamiltonian. Solution of the IB model through the linked cluster expansion~\cite{Mahan00} then enables an efficient analytic calculation of the cumulative effect of all the component parts of the Trotter decomposition.

\subsection{Polaron master equation: Adiabatic CWE}
\label{sec:WR}
We define a Hamiltonian $\mathcal{H}_{\rm CWE}(t)$, which includes both the time-independent Hamiltonian $\mathcal{H}$ [\Eq{eq:Hlab}] and the CWE $\mathcal{V}_{\rm CWE}(t)$ [\Eq{eq:VCW}],
\begin{align}
\mathcal{H}_{\rm CWE}(t \ge t_0) &= \mathcal{H} + \Omega_X (d^{\dagger} e^{-i\omega t} + d e^{i\omega t})\nonumber\\
&\hspace{0.5cm}+ \Omega_C (a^{\dagger} e^{-i\omega t} + a e^{i\omega t})\,,
\end{align}
where we have allowed for excitation through both exciton and cavity channels, each with their own excitation strength $\Omega_{X,C}$. To calculate the steady-state absorption, we impose the condition $t \to \infty$ on the expression for absorption \Eq{eq:abs_CWE2},
\begin{equation}
A(\omega) = - \lim_{t \to \infty} \frac{d \rho_{00}(t)}{dt} \,. \label{eq:abs_inf}
\end{equation}

Following standard procedure~\cite{AlsingPRA92,Wilson-RaePRB02,NazirJPCM16,McCutcheonNJP10}, we perform a rotating wave transformation on the full time-dependent Hamiltonian $\mathcal{H}_{\rm CWE}(t)$ to remove the time-dependent terms $e^{\pm i \omega t}$,
\begin{equation}
\mathcal{H}_{\rm CWE} \to \mathcal{H}_{\rm CWE}' = \hat{Y}(t) \mathcal{H}_{\rm CWE} \hat{Y}^{\dagger}(t) - i \hat{Y}(t) \frac{d \hat{Y}^{\dagger}(t)}{dt}\,,
\end{equation}
with transformation operator $\hat{Y}(t) = e^{i (d^{\dagger}d + a^{\dagger}a) \omega t}$.

We focus on the regime of strong exciton-phonon interactions, in which the {\it polaron} master equation is most appropriate. A unitary transformation, known as the {\it polaron transformation}~\cite{Wilson-RaePRB02,NazirJPCM16,McCutcheonNJP10,HorneckerPRB17}, is performed on the system Hamiltonian $\mathcal{H}_{\rm CWE}'$,
\begin{equation}
\mathcal{H}_{\rm CWE}' \to \mathcal{H}_{\rm CWE}'' = e^{\hat{S}} \, \mathcal{H}_{\rm CWE}' \, e^{-\hat{S}}\,,\label{eq:poltransdef}
\end{equation}
with polaron transformation matrix $\hat{S}$ given by
\begin{equation}
\hat{S} = d^{\dagger}d \sum_q \left(\frac{\lambda_q}{\omega_q}b^{\dagger}_q - \frac{\lambda_q^*}{\omega_q}b_q\right)\,.\label{eq:S_transform}
\end{equation}
Physically, this transformation acts to ``dress'' the exciton with a phonon cloud, modifying the exciton-cavity subsystem to a polaron-cavity system. The transformed Hamiltonian $\mathcal{H}_{\rm CWE}''$ may be expressed as a sum of three parts:
\begin{equation}
\mathcal{H}_{\rm CWE}'' = \mathcal{H}_{\rm sys} + \mathcal{H}_{\rm ph} + \mathcal{H}_{\rm int}\,, \label{eq:Hprime_3parts}
\end{equation}
representing the {\it system} (polaron-cavity), the {\it bath} (phonon modes) and the {\it interaction} (coupling of the system to the phonon bath) respectively. The three parts of the full polaron-frame Hamiltonian \Eq{eq:Hprime_3parts} have the following forms:
\begin{align}
\mathcal{H}_{\rm sys} &= (\bar{\omega}_X - \omega) d^\dagger d + (\omega_C - \omega) a^\dagger a + \bar{g} (a^\dagger d + d^\dagger a)\nonumber\\
&\hspace{0.5cm}+ \vphantom{\sum_q}\bar{\Omega}_X(d^{\dagger} + d) + \Omega_C(a^{\dagger} + a) \,,\label{eq:Hprime_sys}\\
\mathcal{H}_{\rm int} &= \sum_{\alpha = g,u} X_{\alpha} \otimes B_{\alpha} \,,\label{eq:Hprime_int}
\end{align}
where $\bar{\omega}_X = \omega_X + \Omega_p$ is the exciton frequency $\omega_X$ modified by the polaron shift $\Omega_p$, $\bar{g} = g \langle B \rangle$ ($\bar{\Omega}_X = \Omega_X \langle B \rangle$) is the phonon-renormalized exciton-cavity coupling (excitation) strength. Throughout this paper, the bar notation [$\bar{\omega}_X$, for example] will be used to denote the polaron renormalized system. Other parameters within Eqs. (\ref{eq:Hprime_sys}) and (\ref{eq:Hprime_int}) are defined as follows:
\begin{align}
\Omega_p &= - \sum_q \frac{|\lambda_q|^2}{\omega_q}\,,\label{eq:Omega_p_q}\\
B_{\pm} &= \exp\left(\pm \sum_q \left(\frac{\lambda_q}{\omega_q}b^{\dagger}_q - \frac{\lambda_q^*}{\omega_q}b_q\right)\right)\,,\\
\langle B \rangle &= \langle B_{\pm} \rangle = \exp\left(-\frac{1}{2} \sum_q \left| \frac{\lambda_q}{\omega_q} \right|^2 \left(N_q + \frac{1}{2}\right) \right)\,,\label{eq:Bavg}\\
X_g &= g(a^\dagger d + d^\dagger a) + \Omega_X(d + d^{\dagger})\,,\label{eq:Xg}\\
X_u &= ig(d^{\dagger}a - a^{\dagger}d) + i\Omega_X(d - d^{\dagger})\,,\label{eq:Xu}\\\
B_g &= \nicefrac{1}{2}\left(B_+ + B_- - 2\langle B \rangle \right)\,,\label{eq:Bg}\\
B_u &= \nicefrac{i}{2} \left(B_- - B_+\right)\label{eq:Bu}\,,
\end{align}
where $N_q = 1/(e^{\beta \omega_q} - 1)$ is the Bose occupation numbers for a phonon state with energy $\omega_q$.

The polaron shift $\Omega_p$ and thermal phonon expectation value $\langle B \rangle$ may be alternatively expressed as
\begin{align}
\Omega_p &= - \int_0^{\infty} d\omega \frac{J(\omega)}{\omega}\,,\label{eq:omegaP_me}\\
\langle B \rangle &= \exp\left(\frac{1}{2}\int_0^{\infty} d\omega \frac{J(\omega)}{\omega^2}\coth\left(\frac{\beta \omega}{2}\right)\right)\,,\label{eq:B_expec}
\end{align}
where we have converted the summation over $q$ to an integration $\sum_q \rightarrow \frac{\Vol}{(2\pi)^3 v_s^3} \int d^3 \omega$ [where $\Vol$ is the sample volume] and expressed $|\lambda_q|^2$ in terms of the spectral density function $J(\omega) = \sum_{q}|\lambda_q|^2 \delta(\omega - \omega_q)$.

We now look to solve the Lindblad master equation, \Eq{eq:Lind_master}, in the polaron frame. In order to achieve a tractable form of this equation we apply the following assumptions and approximations:
\begin{enumerate}
\item We assume that the interaction term $\mathcal{H}_{\rm int}$ is weak, enabling treatment of this term as a perturbation. Following convention~\cite{Wilson-RaePRB02,McCutcheonNJP10,NazirJPCM16}, we treat the interaction term to second order -- known as the {\it second order Born approximation}.
\item We assume that the phonon bath is sufficiently large to be unaffected by its interaction with the system, thereby enabling factorization of the polaron frame density matrix $\rho(t)$ at all times,
\begin{equation}
\rho(t) = \rho_{S'}(t) \otimes \rho_B\,,
\end{equation}
where $\rho_{S'}$ ($\rho_B$) pertains to the polaron-cavity system (phonon bath), and we take $\rho_B$ as approximately time-independent. In the linear regime, the density matrix of the polaron-cavity system $\rho_{S'}$ may be expressed in a reduced basis of just three states [as shown in \Eq{eq:rhos} in relation to the exciton-cavity density matrix $\rho_S$].
\end{enumerate}
With these simplifying conditions, we arrive at the following form of the Lindblad master equation in the polaron frame:
\begin{align}
&\frac{d \rho_{S'}(t)}{d t} = -i[\mathcal{H}_{\rm sys},\rho_{S'}(t)] + \bar{\mathcal{D}}_S(t)- \int_{t_0}^t d t' \sum_{\alpha = g,u} \nonumber\\
& \times \Big\{ G_{\alpha}(t') \left[X_{\alpha}, \, e^{-i\mathcal{H}_{\rm sys} t'} X_{\alpha} \, \rho_{S'}(t-t') e^{i \mathcal{H}_{\rm sys} t'}\right]+ \text{H.c.}\Big\}\,, \label{eq:Lindpolaron}
\end{align}
see \cite{Wilson-RaePRB02} and \App{app:2ndborn} for derivation.
Here, ``H.c.'' denotes the Hermitian conjugate of all preceding terms within the braces $\{\ldots\}$. In arriving at \Eq{eq:Lindpolaron}, we have exploited the factorization of system and bath Hilbert spaces within $\mathcal{H}_{\rm int}$ and introduced the polaron Greens functions $G_{\alpha}(t)$~\cite{Wilson-RaePRB02},
\begin{equation}
G_{\alpha}(t) = \langle B_{\alpha} \, e^{-i \mathcal{H}_{\rm ph} t}B_{\alpha}e^{i \mathcal{H}_{\rm ph} t}\rangle 
\label{eq:Gal}
\end{equation}
with $\langle\dots\rangle$ denoting the expectation value taken over the states of the phonon system in equilibrium.
Explicitly, $G_{g}$ and $G_u$ have the following forms:
\begin{align}
G_g(t) &= \langle B \rangle^2 [\cosh\phi(t) - 1]\,\label{eq:Gg_t}\\
G_u(t) &= \langle B \rangle^2 \sinh \phi(t)\,,\label{eq:Gu_t}
\end{align}
with $\phi(t)$ defined in terms of the phonon spectral density $J(\omega)$ as
\begin{equation}
\phi(t) = \int_0^{\infty} d\omega \frac{J(\omega)}{\omega^2} \left(\coth\frac{\beta \omega}{2}\cos (\omega t) - i\sin(\omega t) \right)\,.\label{eq:phi_t}
\end{equation}

From \Eq{eq:cf_full}, we note that polaron-cavity density matrix element $\rho_{00}(t) = a_0(t) a_0^*(t)$ is zeroth order in perturbation strength $\Omega_{X,C}$, whilst $\rho_{X0}(t)$ and $\rho_{C0}(t)$ [alongside their respective Hermitian conjugates] are both first order. If we neglect second order and beyond within the product $X_{\alpha}\rho_{S'}(t-t')$, the commutator in \Eq{eq:Lindpolaron} becomes
\begin{align}
\left[X_{\alpha}, \, e^{-i\mathcal{H}_{\rm sys} t'} X_{\alpha} \, \rho_{S'}(t-t') e^{i \mathcal{H}_{\rm sys} t'}\right] &=\nonumber\\
&\hspace{-3cm} X_{\alpha} \,e^{-i\left(\bar{\mathcal{H}}_{\rm JC} - \mathbb{1}\omega\right)t'} X_{\alpha} \,\rho_{S'}(t-t')\,,
\end{align}
where $\mathbb{1}$ is the $2 \times 2$ identity matrix and $\bar{\mathcal{H}}_{\rm JC}$ is the polaron-transformed Jaynes-Cummings Hamiltonian,
\begin{equation}
\bar{\mathcal{H}}_{\rm JC} = \bar{\omega}_X d^{\dagger}d + \omega_C a^{\dagger}a + \bar{g}(a^{\dagger}d + d^{\dagger}a)\,. \label{eq:HJC_pol}
\end{equation}
We now define a reduced polaron-cavity subsystem density matrix,
\begin{equation}
R(t) = \sum_{m=0,X,C} \rho_{m0}(t) \ket{m}\bra{0}\,.
\end{equation}
According to \Eq{eq:Lindpolaron}, its evolution is described by the following reduced master equation:
\begin{equation}
\frac{d R(t)}{dt} =  -i H_{\rm sys} \, R(t) 
- \int_{0}^{t} d t' \, e^{i\omega t'}\,\hat{M}(t')\,R(t-t') + \text{H.c.}\,,
\label{eq:rhodot_CWE}
\end{equation}
where $3 \times 3$ matrices $H_{\rm sys}$ and  $\hat{M}(t)$ are defined as follows:
\begin{align}
H_{\rm sys} &= \mathcal{H}_{\rm sys} - i\gamma_X d^{\dagger}d - i\gamma_C a^{\dagger}a \,,\label{eq:Hsys_CW_comp}\\
\hat{M}(t) &= \begin{pmatrix} \Omega_X^2 W_{XX}(t) & g\Omega_X W_{XC}(t) & g \Omega_X W_{XX}(t)\\
g \Omega_X W_{CX}(t) & g^2 W_{XX}(t) & g^2 W_{XC}(t)\\
g \Omega_X W_{XX}(t) & g^2 W_{CX}(t) & g^2 W_{CC}(t)\end{pmatrix}\label{eq:MCW}
\end{align}
with 
\begin{align}
W_{jk}(t) &= \begin{cases}
U_{jk}(t) \, G_+(t) & \text{for}\, j=k\,,\\
U_{jk}(t) \, G_-(t) & \text{for}\, j\neq k\,,\\
\end{cases}\label{eq:Wjk_t}\\
U_{jk}(t) &= \bra{j} e^{-i \bar{\mathcal{H}}_{\rm JC} t} \ket{k}\,,\label{eq:Ujk}\\
G_{\pm}(t) &= G_g(t) \pm G_u(t) = \langle B \rangle^2 \left(e^{\pm \phi(t)} -1\right)\,. \label{eq:Gpm}
\end{align}
Note that $H_{\rm sys}$ is a complex Hamiltonian; the additional (imaginary) terms relative to $\mathcal{H}_{\rm sys}$ are equivalent to those contained within the dissipator $\bar{\mathcal{D}}_S(t)$ of \Eq{eq:Lindpolaron}.


To zeroth perturbation order $\rho_{00}=1$, and hence transitions from the exciton and cavity modes (associated with $d\rho_{X0}/dt$ and $d\rho_{C0}/dt$ respectively) are negligible in comparison to transitions from the absolute ground state $d\rho_{00}/dt$. Accordingly, taking the limit $t\to\infty$, \Eq{eq:rhodot_CWE} becomes
\begin{equation}
\lim_{t \to \infty} \begin{pmatrix} d\rho_{00}(t)/dt\\ 0 \\ 0\end{pmatrix} = - \hat{\mathcal{Q}}(\omega) \begin{pmatrix} 1 \\ \rho_{X0}(t \to \infty) \\ \rho_{C0}(t \to \infty) \end{pmatrix} + \text{H.c.}\,.\label{eq:rhodot_CWE_red}
\end{equation}
Matrix $\hat{\mathcal{Q}}(\omega)$ has the form
\begin{equation}
\hat{\mathcal{Q}}(\omega) = i H_{\rm sys} + \begin{pmatrix} \Omega_X^2 \mathcal{W}_{XX} & g\Omega_X \mathcal{W}_{XC} & g\Omega_X \mathcal{W}_{XX} \\ g \Omega_X \mathcal{W}_{CX} & g^2 \mathcal{W}_{XX} & g^2 \mathcal{W}_{XC} \\ g\Omega_X \mathcal{W}_{XX} & g^2\mathcal{W}_{CX} & g^2 \mathcal{W}_{CC}\end{pmatrix}, \label{eq:Qmatrix}
\end{equation}
where $H_{\rm sys}$ is given by \Eq{eq:Hsys_CW_comp} and $\mathcal{W}_{jk}(\omega)$ is the Fourier-Laplace transform of $W_{jk}(t)$,
\begin{equation}
\mathcal{W}_{jk}(\omega) = \int_0^{\infty} dt \, e^{i\omega t}\,W_{jk}(t)\,.
\end{equation}
Note that in deriving \Eq{eq:rhodot_CWE_red} from \Eq{eq:rhodot_CWE} we have used the fact that
\begin{align}
&\lim_{t \to \infty} \int_{0}^{t} d t' \, e^{i\omega t'}\,\hat{M}(t')\,R(t-t') \nonumber\\
&\hspace{2cm}=\int_{0}^{\infty} d t' \, e^{i\omega t'}\,\hat{M}(t') R(t \to \infty)\,,
\label {MRint}
\end{align}
following from the finiteness of the phonon memory time. In fact, $G_\pm(t)\to 0$ for $t\gg \tau_{\rm IB}$ leading to \Eq{MRint}.

From Eqs. (\ref{eq:abs_inf}) and (\ref{eq:rhodot_CWE_red}), we  find the absorption under exciton (cavity) excitation $A_{X(C)}(\omega)$ in the adiabatic limit $\Omega_{C(X)}\to 0$:
\begin{align}
A_X(\omega) &= \Re \left\{\mathcal{W}_{XX} + \vec{f}_X \cdot \hat{\mathcal{Q}}_{R}^{-1} \vec{f}_X\right\}\,, \label{eq:abs_X_CW}\\
A_C(\omega) &= \Re \left\{\vec{f}_C \cdot \hat{\mathcal{Q}}_{R}^{-1} \vec{f}_C\right\}\,, \label{eq:abs_C_CW}
\end{align}
see \App{app:CWE} for intermediate steps. Here, in line with the pulsed excitation regime 
(see \Sec{sec:me_pulsed})
we have dropped the unimportant factor of $2$ and normalized the absorption to excitation strength $\Omega_{X,C}$. Within \Eq{eq:abs_C_CW}, $\hat{\mathcal{Q}}_R$ is a reduced $2 \times 2$ form of matrix $\hat{\mathcal{Q}}$ [\Eq{eq:Qmatrix}], in which only $\ket{X}\bra{X}$, $\ket{X}\bra{C}$, $\ket{C}\bra{X}$ and $\ket{C}\bra{C}$ elements are retained,
\begin{align}
\hat{\mathcal{Q}}_R(\omega) &= i \bar{H}_{\rm JC} - i\omega\mathbb{1} + g^2 \begin{pmatrix} \mathcal{W}_{XX} & \mathcal{W}_{XC} \\ \mathcal{W}_{CX} & \mathcal{W}_{CC} \end{pmatrix}\,,
\label{eq:calQR}
\\
\bar{H}_{\rm JC} &= \begin{pmatrix} \bar{\omega}_X - i\gamma_X &\bar{g} \\ \bar{g} & \omega_C - i\gamma_C \end{pmatrix}\,,\label{eq:QR}
\end{align}
and vectors $\vec{f}_{X,C}$ are given by
\begin{equation}
\vec{f}_X = \begin{pmatrix} \langle B \rangle - ig\mathcal{W}_{CX} \\ - ig\mathcal{W}_{XX} \end{pmatrix}\,, \hspace{0.8cm} \vec{f}_C = \begin{pmatrix} 0 \\ 1 \end{pmatrix}\,.\label{eq:fvector_CWE}
\end{equation}
Note that $\bar{H}_{\rm JC}$ is the complex extension of the polaron transformed Jaynes-Cummings Hamiltonian $\bar{\mathcal{H}}_{\rm JC}$ [\Eq{eq:HJC_pol}], which additionally includes the imaginary dephasing terms $\gamma_{X,C}$.

\subsection{Polaron master equation: Pulsed excitation}
\label{sec:me_pulsed}
We now apply the polaron master equation technique to the case of a pulsed excitation $\mathcal{V}_{\delta}$. In the linear regime, the excitation operator $c_e^{\dagger}$ has the form $\ket{j}\bra{0}$, where $\ket{j} = \ket{X}$ ($\ket{C}$) for excitation in the exciton (cavity) mode. The density matrix immediately after application of the pulsed excitation $\rho(0+)$ [\Eq{eq:rho0+}] may therefore be simplified to
\begin{equation}
\rho(0+) = -i\Theta_e \, \ket{e} \bra{0} \otimes \rho_{\rm ph}\,,\label{eq:rho0+_lin}
\end{equation}
keeping only the linear in $\Theta_e$ terms. Here, we have also taken the exciton-cavity subsystem to be in its absolute ground state prior to excitation, $\rho(-\infty) = \ket{0}\bra{0} \otimes \rho_{\rm ph}$. 

Following a similar procedure to \Sec{sec:WR}, we apply a polaron transformation to Hamiltonian $\mathcal{H}$ and solve the Lindblad master equation in the polaron frame,
\begin{equation}
\mathcal{H} \to \mathcal{H}' = e^{\hat{S}} \, \mathcal{H} \, e^{-\hat{S}}\,,\label{eq:poltransdef}
\end{equation}
with polaron transformation matrix $\hat{S}$ given by \Eq{eq:S_transform}. As before, the transformed Hamiltonian $\mathcal{H}'$ may be expressed as a sum of three parts, $\mathcal{H}' = \mathcal{H}_{\rm sys}^{\delta} + \mathcal{H}_{\rm ph} + \mathcal{H}_{\rm int}^{\delta}$, where
\begin{align}
\mathcal{H}_{\rm sys}^{\delta} &= \bar{\mathcal{H}}_{\rm JC}\,,\label{eq:Hprime_sys_delta}\\
\mathcal{H}_{\rm int}^{\delta} &= \sum_{\alpha = g,u} X_{\alpha}^{\delta} \otimes B_{\alpha} \,,\label{eq:Hprime_int_delta}\\
X_g^{\delta} &= g(a^{\dagger}d + d^{\dagger}a)\,,\\
X_u^{\delta} &= ig(d^{\dagger}a - a^{\dagger}d)\,,
\end{align}
with $\bar{\mathcal{H}}_{\rm JC}$ given by \Eq{eq:HJC_pol} and $B_{g,u}$ given by Eqs. (\ref{eq:Bg}) and (\ref{eq:Bu}). Note that $\mathcal{H}_{\rm sys}^{\delta}$ and $\mathcal{H}_{\rm int}^{\delta}$ are analogous to their CWE counterparts [Eqs. (\ref{eq:Hprime_sys}) and (\ref{eq:Hprime_int})] but with $\Omega_X$, $\Omega_C$, and $\omega$ set to zero.

Applying the same simplifying conditions as in the CWE case, we arrive at \Eq{eq:Lindpolaron} with $\mathcal{H}_{\rm sys} \to \bar{\mathcal{H}}_{\rm JC}$ and $X_{g,u} \to X_{g,u}^{\delta}$,
\begin{align}
&\frac{d \rho_{S'}(t)}{d t} = -i[\bar{\mathcal{H}}_{\rm JC},\rho_{S'}(t)] + \bar{\mathcal{D}}_S(t)- \int_{t_0}^t d t' \sum_{\alpha = g,u} \nonumber\\
&\times\Big\{ G_{\alpha}(t') \left[X_{\alpha}^{\delta}, \, e^{-i\bar{\mathcal{H}}_{\rm JC} t'} X_{\alpha}^{\delta} \, \rho_{S'}(t-t') e^{i \bar{\mathcal{H}}_{\rm JC} t'}\right]+ \text{H.c.}\Big\}\,, \label{eq:Lindpolaron_pulsed}
\end{align}

In order to address the same quantity (the linear polarization and absorption) we must consider, in contrast to the CWE case, the time-evolution of the reduced density matrix $R^{\delta}(t)$ at all times $t>t_0$, not just in the limit $t \to \infty$. We outline two alternative approaches, NZ and TCL, to solving \Eq{eq:Lindpolaron_pulsed} in the following subsections.

\subsubsection{NZ master equation\label{sec:NZ}}
For the case of a pulsed excitation, we find absorption $A(\omega)$ from density matrix elements $\rho_{X0}(t)$ or $\rho_{C0}(t)$, depending upon the mode of excitation and observation [see Eqs. (\ref{eq:genericP}) and (\ref{eq:Absorp_pulsed})]. The polaron-cavity density matrix element $\rho_{00}(t)$ is fully decoupled from these elements, enabling us to define a two-basis reduced density matrix,
\begin{equation}
R^{\delta}(t) = \begin{pmatrix} \rho_{X0}(t) \\ \rho_{C0}(t) \end{pmatrix}\,.\label{eq:Rdelta}
\end{equation}
Assuming, without loss of generality, that the pulsed excitation is applied at time $t_0=0$, this reduced density matrix obeys a relation equivalent to \Eq{eq:rhodot_CWE},
\begin{equation}
\frac{d R^{\delta}(t)}{dt} = -i \bar{H}_{\rm JC} R^{\delta}(t)
+ \int_{0}^t d t' \, \hat{M}^{\delta}(t') R^{\delta}(t-t')\,,
\label{eq:rhodot_pulsed}
\end{equation}
where $\bar{H}_{\rm JC}$ and $\hat{M}^{\delta}(t)$ are $2 \times 2$ analogues of $H_{\rm sys}$ and $\hat{M}(t)$, given by \Eqs{eq:Hsys_CW_comp}{eq:MCW}, respectively:
\begin{align}
\bar{H}_{\rm JC} &= \begin{pmatrix} \bar{\omega}_X - i\gamma_X & \bar{g} \\ \bar{g} & \omega_C - i\gamma_C \end{pmatrix}\,,\\
\hat{M}^{\delta}(t) &= g^2 \begin{pmatrix} W_{XX}(t) & W_{XC}(t) \\ W_{CX}(t) & W_{CC}(t) \end{pmatrix}\,.
\end{align}

Noting that the integral term in \Eq{eq:rhodot_pulsed} describes a convolution, it is natural to explore the viability of solving the equation in Fourier space. Applying a Fourier-Laplace transformation to \Eq{eq:rhodot_pulsed}, we obtain
\begin{equation}
\mathcal{R}(\omega) = \hat{\mathcal{Q}}_R^{-1} \, R^{\delta}(t=0_+)\,, \label{eq:sigma_omega}
\end{equation}
with $\hat{\mathcal{Q}}_R$ defined in \Eq{eq:QR} and $R^{\delta}(t=0_+)$ found from \Eq{eq:rho0+_lin}. 

Clearly, the reduced density matrix \Eq{eq:Rdelta} represents the linear polarization. Using the link between the linear polarization and absorption, given by \Eq{eq:Absorp_pulsed}, we find the absorption under exciton (cavity) mode excitation $A_{X(C)}(\omega)$:
\begin{equation}
A_j(\omega) = \Re \left\{ \vec{F}_j\cdot \, \hat{\mathcal{Q}}_R^{-1} \vec{F}_j \right\}\,, \label{eq:abs_j_NZ}
\end{equation}
with $j=X,C$, and $\vec{F}_{j}$ given by
\begin{equation}
\vec{F}_X = \begin{pmatrix} \langle B \rangle \\ 0 \end{pmatrix}\,, \hspace{0.8cm} \vec{F}_C = \begin{pmatrix} 0 \\ 1 \end{pmatrix}\,. \label{eq:fvector_pulsed}
\end{equation}
Note that in arriving at Eqs. (\ref{eq:abs_j_NZ}) and (\ref{eq:fvector_pulsed}), we have accounted for the transformation of excitation and observation operators $c_{e,o}$ into the polaron frame, which contributes a factor of $\langle B \rangle^2$ to $A_X(\omega)$.

It is instructive to compare Eqs. (\ref{eq:abs_j_NZ}) and (\ref{eq:fvector_pulsed}) to the equivalent expressions for the case of an adiabatic CWE, Eqs. (\ref{eq:abs_X_CW})--(\ref{eq:fvector_CWE}). Clearly, both results are expressed in terms of the same kernel function $\hat{\mathcal{Q}}_R(\omega)$. The expression for absorption in the cavity mode $A_C(\omega)$ in the present (pulsed) regime is identical to that for adiabatic CWE. This, however, is not true when feeding is via the excitonic mode. Considering the latter case in greater detail, the excitonic absorption $A_X(\omega)$ in the pulsed regime [\Eq{eq:abs_j_NZ}] may be expressed as
$A_X(\omega) = {\rm Re}\{\langle B \rangle^2 (\hat{\mathcal{Q}}_R^{-1})_{11}\}$; 
whilst this term appears in the expression for absorption $A_X(\omega)$ in the adiabatic CWE regime
[\Eq{eq:abs_X_CW}], it is modified by the addition of several other terms. 


\subsubsection{TCL master equation\label{sec:TCL}}
A widely employed alternative approach to solving \Eq{eq:Lindpolaron_pulsed} is the time-convolutionless master equation. In its present form, \Eq{eq:Lindpolaron} has memory: the future evolution of the polaron-cavity subsystem density matrix $\rho_{S'}(t)$ depends upon its history through the term $\rho_{S'}(t - t')$. In the TCL approach, we remove this memory through the {\it Markov approximation}, in which we make the replacement
\begin{equation}
\tilde {\rho}_{S'}(t-t')  \to \tilde{\rho}_{S'}(t)\,,
\end{equation}
The Markov approximation is valid if the exciton-cavity timescale $\tau_{\rm JC}$ is large in comparison to the bath memory time $\tau_{\rm IB}$. The resulting time-local equation has the form
\begin{align}
&\frac{d \rho_{S'}(t)}{d t} = -i[\mathcal{H}_{\rm sys}^{\delta},\rho_{S'}(t)] + \bar{\mathcal{D}}_S(t)
\nonumber\\
&- \int_{t_0}^t d t' \sum_{\alpha = g,u} \Big\{G_{\alpha}(t') \left[X_{\alpha}^{\delta}, \, \tilde{X}_{\alpha}^{\delta}(t') \rho_{S'}(t) \right]
+ \text{H.c.}\Big\}\,. \label{eq:Lindpolaron_TCL}
\end{align}
where tilde denotes the interaction representation of operators, $\tilde{X}(t)=e^{i \mathcal{H}_{\rm sys}^{\delta} t} X e^{-i \mathcal{H}_{\rm sys}^{\delta} t}$.

As  in the NZ case, we may limit our consideration to only certain elements of the polaron-cavity density matrix $\rho_{S'}(t)$. Taking the same reduced density matrix $R^{\delta}(t)$ as for the NZ approach [\Eq{eq:Rdelta}], \Eq{eq:Lindpolaron_TCL} is simplified to
\begin{equation}
\frac{d R^{\delta}(t)}{d t} = - \hat{Q}_{\rm TCL}(t) R^{\delta}(t)\,, \label{eq:dsigma_TCL}
\end{equation}
where matrix $\hat{Q}_{\rm TCL}(t)$ is given by
\begin{equation}
\hat{Q}_{\rm TCL}(t) = i\bar{H}_{\rm JC} + g^2 \int_{t_0}^t dt' \, \sum_{+,-} G_{\pm}(t') \hat{M}_{\pm}(t')\,, \label{eq:QTCL}
\end{equation}
with $G_{\pm}(t)$ defined in\Eq{eq:Gpm} and $\hat{M}_{\pm}(t)$ described in terms of elements of the polaron-transformed Jaynes-Cummings matrix exponential $U_{jk}(\pm t)$ [\Eq{eq:Ujk}]:
\begin{align}
\hat{M}_{+}(t) &= \begin{pmatrix} U_{CX}(-t)U_{CX}(t) & U_{CX}(-t)U_{CC}(t) \\ U_{XC}(-t)U_{XX}(t) & U_{XC}(-t)U_{XC}(t)\end{pmatrix}\,,\\
\hat{M}_{-}(t) &= \begin{pmatrix} U_{CC}(-t)U_{XX}(t) & U_{CC}(-t)U_{XC}(t) \\ U_{XX}(-t)U_{CX}(t) & U_{XX}(-t)U_{CC}(t)\end{pmatrix}\,.
\end{align}
Note that in contrast to matrix $\hat{\mathcal{Q}}_R(\omega)$ [\Eq{eq:QR}], $\hat{Q}_{\rm TCL}(t)$ is defined in the time domain and thus dictates the temporal evolution of the reduced density matrix $R^{\delta}(t)$. This approach therefore enables calculation of the polarization $P(t)$ according to \Eq{eq:genericP}. We then take the real part of the inverse Fourier transform of polarization, as outlined in \Eq{eq:Absorp_pulsed}, to determine the absorption $A(\omega)$. Since the kernel matrix $\hat{Q}_{\rm TCL}(t)$ in the TCL master equation (\ref{eq:dsigma_TCL}) does not have a convolution form, the TCL result is purely numerical and does not allow any qualitative comparison with the NZ and CWE solutions both having analytic form. However, we provide a quantitative comparison below. 


\subsection{Trotter decomposition with cumulant expansion for pulsed excitation}
\label{sec:Trotter}
The underlying principle of the Trotter decomposition method is the separation of the full system Hamiltonian $\mathcal{H}$ [\Eq{eq:Hlab}] into two parts, associated with the JC and IB models respectively,
\begin{align}
\mathcal{H} &= \mathcal{H}_{\rm JC} + \mathcal{H}_{\rm IB}\,\\
\mathcal{H}_{\rm JC} &= \omega_X d^{\dagger}d + \omega_C a^{\dagger}a + g(a^{\dagger}d + d^{\dagger}a)\,\\
\mathcal{H}_{\rm IB} &= \mathcal{H}_{\rm ph} + d^{\dagger}d V\,,
\end{align}
with $\mathcal{H}_{\rm ph}$ and $V$ defined in \Eq{eq:HIBcomponents}. Each of these parts is individually exactly solvable, yet the combination of the two exact solutions to determine the dynamics of the full system presents a significant challenge. A solution proposed by the present authors~\cite{MorreauPRB19} utilizes the linked cluster expansion solution to the IB model to enable effective combination of JC and IB components.

To accommodate dissipation within the Trotter decomposition method, we define a {\it non-Hermitian} Hamiltonian $H$, which accounts for the influence of the radiative decay $\gamma_C$ and long-time ZPL dephasing $\gamma_X$ without the need for a dissipator term $\mathcal{D}$,
\begin{equation}
H = \mathcal{H} - i\gamma_X d^{\dagger} d - i\gamma_C a^{\dagger}a\,,
\end{equation}
where the Hermitian Hamiltonian $\mathcal{H}$ is defined in \Eq{eq:Hlab}. We may re-express the Lindblad master equation [\Eq{eq:Lind_master}] in terms of the complex Hamiltonian $H$,
\be
i\dot{\rho}=H\rho-\rho H^\ast +2i\gamma_Xd\rho d^\dagger+2i\gamma_C a\rho a^\dagger\,.
\label{eq:Lindblad_comp}
\ee
Considering only linear polarization, we reduce the density matrix to only $\ket{X} \bra{0}$ and $\ket{C} \bra{0}$ elements. The third and fourth terms on the RHS of \Eq{eq:Lindblad_comp} therefore vanish, yielding an explicit solution:
\be
\rho(t)=e^{-iHt}\rho(0_+)e^{iH^\ast t}\,,\label{eq:rhot_comp}
\ee
with the density matrix immediately after excitation $\rho(0_+)$ given to the linear approximation by \Eq{eq:rho0+_lin}. Applying Trotter's decomposition theorem to separate complex Hamiltonian $H$ into JC and IB parts, we may express \Eq{eq:rhot_comp} as
\begin{align}
\rho(t) &= \lim_{N \to \infty} \left( \mathcal{T} \prod_{n=1}^N e^{-iH_{\rm IB}(t_n - t_{n-1})} e^{-i\mathcal{H}_{\rm JC}(t_n - t_{n-1})} \right)  \nonumber\\
&\times \rho(0_+) \left( \mathcal{T}_{inv} \prod_{n=1}^N e^{iH_{\rm IB}(t_n - t_{n-1})} e^{i\mathcal{H}^*_{\rm JC}(t_n - t_{n-1})} \right)\,, \label{eq:rho_t_trotter}
\end{align}
where $\mathcal{T}$ and $\mathcal{T}_{inv}$ are, respectively the time ordering and inverse time ordering operators.

Converting the density matrix to the interaction representation ${\rho}_I(t) = e^{i\mathcal{H}_{\rm ph}t} \rho(t) e^{-i\mathcal{H}_{\rm ph}t}$ and exploiting the commutativity of $H_{\rm ph}$ and $\mathcal{H}_{\rm JC}$, \Eq{eq:rho_t_trotter} becomes
\begin{align}
{\rho}_I(t) &= \lim_{N \to \infty} \left(\mathcal{T} \prod_{n=1}^{N} \hat{W}(t_n,t_{n-1}) \hat{M}(t_n - t_{n-1})\right) \nonumber\\
&\times \rho(0_+) \left(\mathcal{T}_{inv} \prod_{n=1}^N \hat{W}^{\dagger}(t_n,t_{n-1}) \hat{M}^{\dagger}(t_n - t_{n-1}) \right)\,,\label{eq:rho_t_int}
\end{align}
where we have introduced two new operators, $\hat{M}$ and $\hat{W}$, associated with the JC and IB Hamiltonians, respectively,
\begin{align}
\hat{M}(t_n - t_{n-1}) &= \hat{M}(\Delta t) = e^{-i H_{\text{JC}} \Delta t} \label{eq:M},\\
\hat{W}(t_n,t_{n-1}) &= e^{i H_{ph} t_n}e^{-i \mathcal{H}_{\text{IB}} \Delta t}e^{-i \mathcal{H}_{\rm ph} t_{n-1}},
\end{align}
with $\Delta t = t/N$. Both operators $\hat{M}$ and $\hat{W}$ may be expressed as $2\!\times \!2$ matrices in the $\ket{X}$, $\ket{C}$ basis. Operator $\hat{W}$, associated with the IB Hamiltonian, has the form
\be
\hat{W}(t_n,t_{n-1})=\begin{pmatrix} W_X(t_n,t_{n-1}) &0\\0 & W_C(t_n,t_{n-1})\end{pmatrix}
\ee
with
\bea
W_X(t_n,t_{n-1}) &=& e^{iH_{\rm ph}t_n} e^{-i (H_{\rm ph}+V) (t_n-t_{n-1})} e^{-iH_{\rm ph}t_{n-1}}\,,\nonumber\\
W_C(t_n,t_{n-1}) &=& 1\,.
\label{Woper}
\eea

To find the linear optical polarization from the density matrix ${\rho}_I(t)$ we apply \Eq{eq:genericP} in the interaction representation defined above. Keeping in mind the aim of calculating absorption, we select the observation mode to match the excitation mode, $c_o = c_e = \ket{0}\bra{j}$.

Substituting for $\rho(0_+)$ from \Eq{eq:rho0+_lin}, all $\hat{W}^{\dagger}$ and $\hat{M}^{\dagger}$ matrices to the right of $\rho(0_+)$ act on state $\bra{0}$ and hence reduce to unity. Hence, dropping the unimportant factors of $i$ and $\Omega_e$, we arrive at the following expression for linear polarization:
\begin{equation}
P_{jj}(t) = \left\langle \bra{j} \mathcal{T} \prod_{n=1}^{N} \hat{W}(t_n,t_{n-1})\hat{M}(t_n - t_{n-1}) \ket{j} \right\rangle \label{eq:Pdef}
\end{equation}
where $\langle \cdots \rangle$, as in \Eq{eq:Gal}, denotes the trace over all phonon states. Re-expressing the matrix products within \Eq{eq:Pdef} explicitly as summations over individual matrix elements, we arrive at the following expression for polarization,
\begin{align}
P_{jj}(t) &= \sum_{i_{N-1} = X,C} \ldots \sum_{i_1=X,C} M_{i_N i_{N-1}} \ldots M_{i_2 i_1} M_{i_1 i_0} \nonumber \\
& \times \left\langle W_{i_N}(t,t_{N-1}) \ldots W_{i_2}(t_2,t_1)W_{i_1}(t_1,0)\right\rangle\,, \label{eq:Pjk}
\end{align}
where $i_N = i_0 = j$, and we have exploited the $2 \times 2$ matrix form of the JC operator $\hat{M}$: $M_{i_n i_m} = [\hat{M}(\Delta t)]_{i_n i_m}$.

We describe a particular set of $i_n$ as a {\it permutation}. Consider, for example, the permutation $i_1=X$, $i_2=X$, $i_3=C$ etc. Physically, this corresponds to the exciton-phonon subsystem being in the exciton state $\ket{X}$ in the first time interval, $t_1$, remaining in this state for the second time interval, $t_2 - t_1$, and then transferring to the cavity state $\ket{C}$ at the time moment $t_2$ and staying in the cavity state during the third time interval, $t_3 - t_2$. This reversible transfer between exciton and cavity states is a fundamental property of the strong coupling regime. 

Evaluation of \Eq{eq:Pjk} requires the summation over all possible permutations. This summation has been historically achieved through somewhat obtuse algorithms comprising complex path selection rules~\cite{NahriJPCM16,GlasslPRB12,VagovPRL14,VagovPRL07}. An earlier work by the present authors~\cite{MorreauPRB19} presents a linked cluster expansion technique~\cite{Mahan00}. Crucially, this technique enables the over all permutations summation to be performed in the exponent and thus translated into a computationally simple matrix product.

As part of the linked cluster expansion solution, we associate with each permutation a step-function $\hat{\theta}(\tau)$ being equal to 0 (1) over the time interval $t_n - t_{n-1}$ if $i_n = C$ ($i_n = X$) [the system is in state $\ket{C}$ ($\ket{X}$)]~\cite{MorreauPRB19}. The product of $W$-operators for a particular permutation can be written as
\begin{equation}
W_{i_N}(t,t_{N-1}) \ldots W_{i_1}(t_1,0) = \mathcal{T} \exp{-i\int_{0}^{t} \bar{V}(t') d t'}\,, \label{eq:productW}
\end{equation}
where $\bar{V}(t') = \hat{\theta}(t'){V}_I(t')$. Calculating the trace of \Eq{eq:productW} over all phonon states we obtain
\begin{equation}
\left \langle W_{i_N}(t,t_{N-1}) \ldots W_{i_2}(t_2,t_1)W_{i_1}(t_1,0) \right\rangle_{B} = e^{\bar{K}(t)}, \label{eq:expWlincum}
\end{equation}
where
\begin{equation}
\bar{K}(t) = - \frac{1}{2} \int_0^t d t' \int_0^t d t'' \langle \mathcal{T} \bar{V}(t') \bar{V}(t'')\rangle
\label{eq:lincum}
\end{equation}
is the linear cumulant for the particular permutation. The explicit dependence of $\bar{K}(t)$ on the specific indices $i_n$ of the permutation is
given by
\begin{equation}
\bar{K}(t) = \sum_{n=1}^N \sum_{m=1}^N \delta_{i_n X}\delta_{i_m X} K_{|n - m|}\,, \label{eq:Ksum}
\end{equation}
where $\delta_{jk}$ is the Kronecker delta and cumulant elements $K_{|n-m|}$ are given by
\begin{equation}
K_{|n - m|} = - \frac{1}{2} \int_{t_{n-1}}^{t_n} d t' \int_{t_{m-1}}^{t_m} d t'' \langle \mathcal{T} V(t')V(t'')\rangle.\label{eq:Knm}
\end{equation}
Note that $K_{|n - m|}$ depends only on the time difference $|t_n - t_m| = |n-m| \Delta t$. Furthermore, all $K_{|n - m|}$ can be efficiently calculated from the standard IB model cumulant $K(t) = \mathcal{T} \exp{-i\int_0^t V(t') d t'}$. This IB model cumulant may be expressed in terms of entities previously defined in relation to the polaron master equation approaches
\begin{equation}
K(t) = \phi(t) - i\Omega_p t - S\,,
\end{equation}
where $\phi(t)$ is defined in \Eq{eq:phi_t}, the polaron shift $\Omega_p$ is defined in Eqs. (\ref{eq:Omega_p_q}) and  (\ref{eq:omegaP_me}), and the Huang-Rhys factor $S$ has the form
\be
S = \int_0^{\infty} d\omega \, \frac{J(\omega)}{\omega^2} \coth{\left(\frac{\omega}{2 k_B T}\right)}\,.
\label{eq:Huang_Rhys}
\ee
Note that $S$ is related to the $\langle B \rangle$ [Eqs. (\ref{eq:Bavg}) and (\ref{eq:B_expec})] as $\langle B \rangle = e^{-S/2}$.

Having in mind an application to semiconductor QDs coupled to bulk acoustic phonons, we use the conditions of the super-Ohmic coupling spectral density and a finite phonon memory time $\tau_{\rm IB}$~\cite{VagovPRL07}. This provides a drastic reduction in the number of terms required in the double summation of \Eq{eq:Ksum}. Indeed, we need to take into account only instances in which $|t_m - t_n| \leqslant \tau_{\text{IB}}$. When selecting $\Delta t$, we must also be mindful of the requirement imposed by the Trotter decomposition method: $\Delta t \rightarrow 0$, which in the present application corresponds to the condition $\Delta t \ll \tau_{\rm JC}$: the time spent in the exciton or cavity mode during the periodic Rabi oscillations must be much longer than the time slice $\Delta t$.

\begin{figure}
\includegraphics{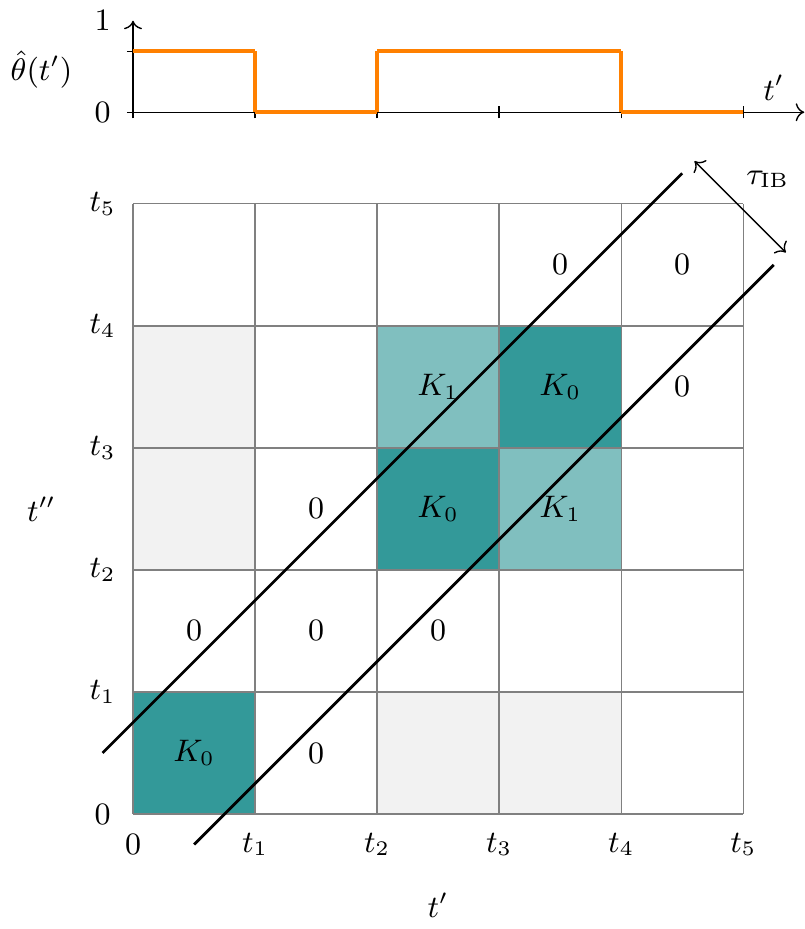}
\caption{Example permutation for the nearest neighbor ($L=2$) implementation with $N=5$. In this instance, $i_1 = X$, $i_2 = C$, $i_3 = X$, $i_4 = X$, $i_5 = C$, as shown in the step function $\hat{\theta}(t)$ shown above the grid.  \label{fig:grid1}}

\hspace{5mm}
%
\includegraphics{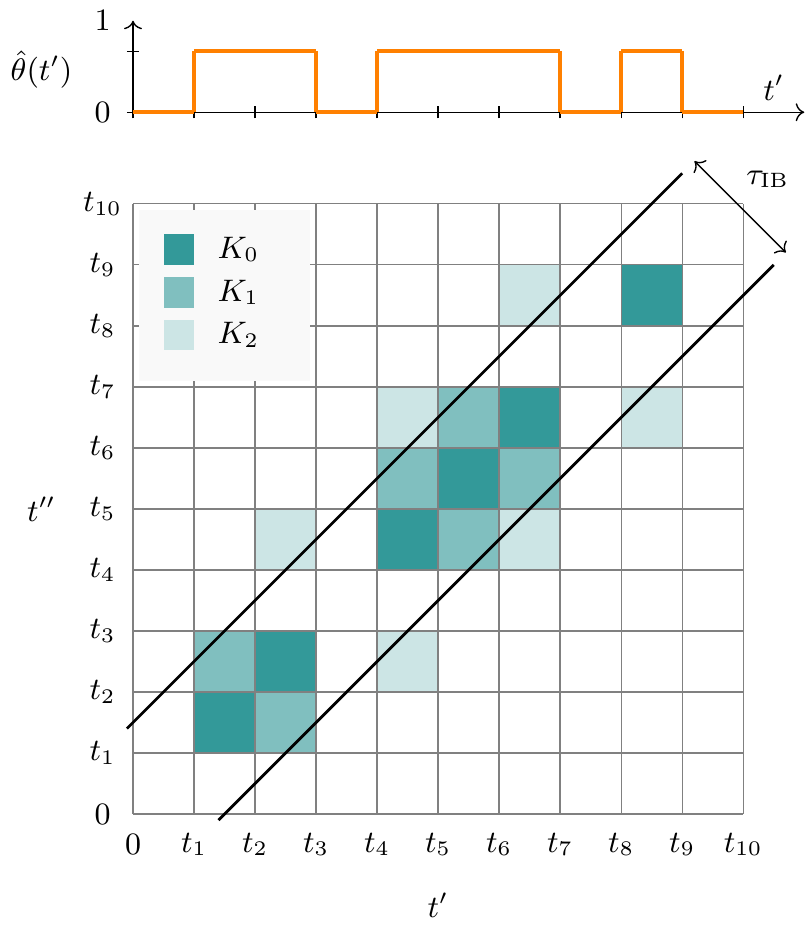}
\caption{Time grid as shown in \Fig{fig:grid1} adjusted for an reduction in the polariton timescale $\tau_{\rm JC}$ by a factor of 2 and depicting a different example permutation. Phonon memory time $\tau_{\rm IB}$ is unchanged and hence an increase in the number of neighbors is required in order to fully account for phonon memory effects. In the present case, we increase $L$ to 3. \label{fig:grid2}}
\label{fig:grid}
\end{figure}

In order to simultaneously satisfy both aforementioned conditions on the time interval $\Delta t$, we must introduce the concept of ``neighbors''. We may graphically depict the double summation of \Eq{eq:Ksum} as a two-dimensional grid such as those shown in \Figs{fig:grid1}{fig:grid}. Choosing $\Delta t$ to satisfy the above-described condition $\Delta t \ll \tau_{\rm JC}$, we then set the number of neighbors L such that the area covered within the double summation of $\bar{K}$ fully encompasses the phonon timescale $\tau_{\rm IB}$. \Fig{fig:grid1} depicts the most straightforward embodiment of the present method, which is suitable for the regime $\tau_{\rm IB} \ll \tau_{\rm JC}$. Here, we consider just nearest neighbors [$L=2$], so that $\bar{K}$ is formed of just two cumulant elements: $K_0$ and $K_1$. Due to the finite phonon memory time $\tau_{\rm IB}$, all other cumulant elements $K_{n>1}$ are vanishing.

It should be noted that the polariton timescale $\tau_{\rm JC}$ is approximately inversely proportional to the exciton-cavity coupling strength $g$, so a greater number of neighbors are required if the exciton-coupling strength $g$ is large. \Fig{fig:grid2} illustrates the case when $\tau_{\rm JC}$ is halved [$g$ approximately doubled] relative to that shown in \Fig{fig:grid1}. If the phonon memory time $\tau_{\rm IB}$ is unchanged, the number of neighbors $L$ must be increased in order to compensate for the reduction in $\tau_{\rm JC}$. In general, we must satisfy the following condition,
\begin{equation}
L \gg \frac{\tau_{\rm IB}}{\tau_{\rm JC}}.
\end{equation}
We focus, in this work, on exciton-cavity coupling strengths in the range \mbox{$50\,\mu$eV $< g < 1.5$ meV}. With other parameters fixed as outlined in \Sec{sec:comparison}, we find $L=15$ to be suitable for all calculations.

Once the appropriate number of neighbors is selected, we define a $2^L$ dimensional matrix $F_{i_L \ldots i_1}^{(n)}$ generated via a recursive relation
\begin{equation}
F_{i_L \ldots i_1}^{(n+1)} = \sum_{l=X,C} G_{i_L \ldots i_1 l} F_{i_{L-1} \ldots i_1 l}^{(n)}\,,
\label{Fs}
\end{equation}
with $G_{i_L \ldots i_1 l}$ given by,
\begin{equation}
G_{i_L \ldots i_1 l} = M_{i_1 l}e^{\delta_{l X}(K_0 + 2\delta_{i_1 X}K_1 \ldots + 2\delta_{i_L X}K_L)}\,.
\label{Gtensor}
\end{equation}
We take $F_{i_L \ldots i_1}^{(1)} = M_{i_1 j}$ as the initial value in the recursive relation, where $\hat{M}$ is defined in \Eq{eq:M}. From \Eq{eq:Pjk}, the polarization is then given by
\begin{equation}
P_{jk}(t)  = e^{\delta_{j X} K_0}  F^{(N)}_{C\ldots C j},
\label{eq:PLneighbours}
\end{equation}
with the absorption determined according to \Eq{eq:Absorp_pulsed}.


\section{Results}


\subsection{Absorption spectra}
\label{sec:comparison}
This aim of this section is to provide a quantitative comparison of the absorption spectra calculated according to the four approaches introduced in \Sec{sec:methods}. To provide a meaningful analysis, we select the following realistic GaAs parameters~\cite{KasprzakNatMat10,AlbertNatComm13}: exciton confinement radius $l = 3.3$\,nm; deformation potential $D_c - D_v = -6.5$\,eV; speed of sound in material $v_s = 4.6 \times 10^3$\,m/s; and, mass density of the material $\rho_m = 5.65$\,g/cm$^3$. The exciton-phonon interaction is fully characterized by the phonon spectral density, which is super-Ohmic for the case of a semiconductor QD coupled to bulk longitudinal acoustic (LA) phonons,
\begin{equation}
J(\omega) = A \, \omega^3 e^{-{\omega^2}/{\omega_0^2}}\,,\label{eq:Jomega}
\end{equation}
found in a spherical model of the QD electron confinement.
Within this expression, $A = (D_c-D_v)^2/ (4 \pi^2 \rho_m v_s^5)$ characterizes the exciton-phonon coupling strength and $\omega_0 = \sqrt{2} v_s/l$ is the cut-off frequency, corresponding to the maximum energy of longitudinal acoustic (LA) phonons~\cite{Krummheuer2002}. The latter is inversely related to the phonon memory time, $\tau_{\text{IB}} \approx 2\pi/\omega_0$, leading to \Eq{eq:tauIB}. Inserting the parameter values outlined above, we find \mbox{$A=0.022 \, {\rm ps}^2$}, \mbox{$\omega_0 = 2.0 \, {\rm ps}^{-1} = 1.3$ meV} and \mbox{$\tau_{\rm IB} = 3.2$ ps}.

\begin{figure}
\includegraphics[width=\columnwidth]{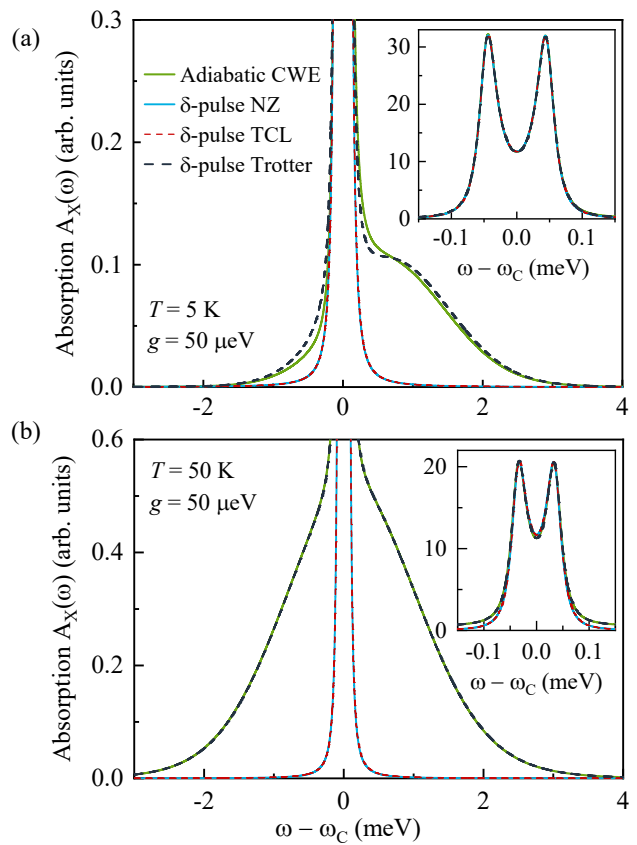}
\caption{(a) Excitonic absorption spectra at $T=5$ K for the case of moderate exciton-cavity coupling strength $g=50\,\mu$eV, calculated according to each of the four above-described methods: adiabatic CWE solved by polaron master equation (shown in solid green); NZ polaron master equation (solid blue); TCL polaron master equation (red short dash); and, Trotter decomposition with linked cluster expansion (black dash). (b) As (a) but at the higher temperature of $T=50$ K. Other parameters reflect realistic InGaAs QDs studied in refs.~\cite{MuljarovPRL04,MuljarovPRL05} and
micropillars studied in refs.~\cite{KasprzakNatMat10,AlbertNatComm13} including $\omega_X=1329.6$\,meV, $\gamma_X = 2\,\mu$eV, $\omega_C=\omega_X + \Omega_p$, $\Omega_p = -49.8\,\mu$eV and $\gamma_C = 30\,\mu$eV.
}
\label{fig:Ax_g50}
\end{figure}

\Fig{fig:Ax_g50}~(a) shows absorption in the excitonic mode $A_X(\omega)$ for the case of moderate exciton-cavity coupling strength $g=50\,\mu$eV at $T=5$ K. The results of each of the three polaron master equation approaches (adiabatic CWE, pulsed NZ and pulsed TCL) are shown, alongside the result of the 15-neighbor Trotter decomposition with linked cluster expansion method. We take the latter as the exact solution, against which to measure the accuracy of the master equation approaches. \Fig{fig:Ax_g50}~(b) shows absorption at the higher temperature of $T=50$ K, with all other parameters unchanged. At both $T=5$ [\Fig{fig:Ax_g50}(a)] and $T=50$ K [\Fig{fig:Ax_g50}(b)], the exact spectrum consists of two distinct peaks superimposed with a phonon broadband (BB). We associate the two peaks with upper and lower polariton lines~\cite{MorreauPRB19}.

Immediately obvious from \Fig{fig:Ax_g50} is the failure of the NZ and TCL methods (introduced in \Sec{sec:me_pulsed}) to capture the phonon BB. When applied to the case of an adiabatic CWE (described in \Sec{sec:WR}), however, the polaron master equation reproduces the broadband to a good degree of accuracy. We may understand this behavior through consideration of the approximations made during the master equation derivation. In particular, all three polaron master equation approaches rely upon the factorization of the full system density matrix $\rho(t)$ into polaron-cavity $\rho_{S'}(t)$ and phononic bath $\rho_B(t)$ Hilbert spaces, see \App{app:2ndborn}.
This decoupling of the exciton-cavity subsystem from the phonon bath, when paired with feeding via a pulsed excitation, masks the true full effect of the phonon environment: the reaction time of the phonon bath $\tau_{\rm IB}$ is simply too slow to adjust to the sudden change in state of the exciton-cavity subsystem. For the case of adiabatic CWE, however, the state of the exciton-cavity subsystem changes on a timescale much longer than $\tau_{\rm IB}$ and hence the phonon bath may react accordingly. 

One may ask why the TD method, which also calculates absorption following a pulsed excitation, is capable of fully capturing the phonon BB. Again, this relates to factorization of the full exciton-cavity-phonon density matrix $\rho(t)$. As discussed in \Sec{sec:Trotter}, the TD method relies upon separation of the density matrix into exciton-cavity and phonon Hilbert spaces. Crucially, however, this factorization is limited to the temporal period prior to and immediately following the pulsed excitation. The delta pulse excitation $V_{\delta}(t)$ [\Eq{eq:pulsed_exc}] acts only on the exciton-cavity subspace and hence the phononic Hilbert space remains in its equilibrium state immediately following the excitation [$t=0_+$]. After this time, the TD method does not factorize the density matrix and hence the response of the phonon bath is fully accounted for.


\begin{figure}
\includegraphics[width=\columnwidth,trim=0 15cm 0 0, clip]{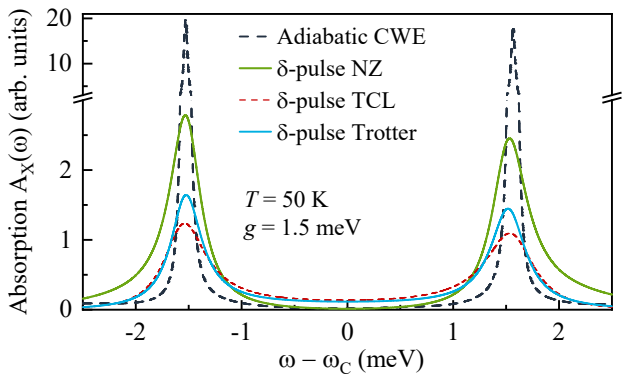}
\caption{As \Fig{fig:Ax_g50}(b) but for the case of very strong exciton-cavity coupling $g = 1.5$ meV.\label{fig:Ax_g1500}
}
\end{figure}

\Fig{fig:Ax_g1500} illustrates absorption in the excitonic mode $A_X(\omega)$ at $T=50$ K and exciton-cavity coupling constant $g=1.5$ meV. Relative to \Fig{fig:Ax_g50}~(b), only the exciton-cavity coupling strength $g$ has been modified. At this substantially stronger coupling strength, we no longer see a distinguishable phonon broadband; the absorption spectrum now consists only of the two polariton peaks. In stark contrast to the findings at $T=5$ K, the results of all three polaron master equation approaches now deviate considerably from the exact (TD) solution.

\subsection{Behavior of polariton line parameters with exciton-cavity coupling strength $g$}
We now seek to quantitively parameterize the two polariton states in order to further explore the accuracy of the polaron master equation techniques. Through the TD method it has been shown~\cite{MorreauPRB19} that each polariton peak is described by a Lorentzian lineshape, giving a full absorption spectrum of the form:
\begin{equation}
A_{X,C}(\omega) = \Re\left\{\sum_{j=1}^2 \frac{C_j}{\omega_j - i\Gamma_j - \omega} + B(\omega)\right\}\,, \label{eq:Apol_lorentz}
\end{equation}
where $B(\omega)$ represents the contribution, if any, from the phonon broadband. The polariton frequencies $\omega_j$ and associated linewidths $\Gamma_j$ are real parameters, whereas the amplitude coefficients $C_j$ may be complex. Note that the parameters within \Eq{eq:Apol_lorentz} may take different values for exciton mode [$A_X(\omega)$] relative to cavity mode [$A_C(\omega)$] excitation; for purposes of brevity, we will consider only the former in the following discussion.

In the TD approach, the parameters $C_{1,2}$, $\omega_{1,2}$ and $\Gamma_{1,2}$ of \Eq{eq:Apol_lorentz} are straightforwardly extracted from the polarization $P_{XX}(t)$ [\Eq{eq:PLneighbours}]. We choose to utilize the polarization $P_{XX}(t)$ rather than the absorption $A_X(\omega)$ for this purpose due to the natural separation of the phonon broadband from the two polariton lines within the time domain: the broadband contributes only to a rapid initial decay of $P_{XX}(t)$, with the remaining long-time asymptotics corresponding to the two polariton lines. A bi-exponential fit, corresponding to the Fourier transform of the Lorentzian part of \Eq{eq:Apol_lorentz}, $\sum_{j=1}^2 C_j (\omega_j - i\Gamma_j - \omega)^{-1}$, is applied to the polarization $P_{XX}(t)$ in the region $t \gg \tau_{\rm IB}$.

The TCL polaron master equation [\Sec{sec:TCL}], like the TD method,  provides a solution in the time domain. The TCL calculation, however, does not capture the phonon broadband and hence the bi-exponential fit may be applied to the full temporal range of the polarization $P_{XX}(t)$.

The NZ and adiabatic CWE polaron master equations [Secs. \ref{sec:NZ} and \ref{sec:WR}, respectively] are solved only in the frequency domain and hence require a different approach in order to extract the parameters of \Eq{eq:Apol_lorentz}.
Separation of the polariton lines is achieved through eigenvalue determination of $\hat{\mathcal{Q}}_R(\omega)$ [\Eq{eq:QR}], with each eigenvalue corresponding to a polariton state. This procedure is best illustrated for the case of zero polaron-cavity detuning $\bar{\omega}_X = \omega_C$ and equal dephasing $\gamma_X = \gamma_C$. In this simple case, outlined in full in \App{app:zerodet}, the eigenvalues of $\hat{\mathcal{Q}}_R$ are given by
\begin{equation}
\lambda_{\pm}(\omega) = -i(\omega - \omega_C  \pm \bar{g}) + g^2\left\{\mathcal{W}_{XX}(\omega) \pm \mathcal{W}_{CX}(\omega)\right\}\,.
\label{eq:lam_pm}
\end{equation}
In the NZ method, the absorption $A_{X}(\omega)$ is simply proportional to the real part of $1/\lambda_{+} + 1/\lambda_{-}$. A similar, albeit slightly more protracted expression characterizes the absorption $A_X(\omega)$ according to the adiabatic CWE master equation [see \Eq{eq:AX_CWE_zerodetuning}]. The polariton lineshapes, as calculated by the NZ and adiabatic CWE master equation approaches, are not Lorentzian in form and hence \Eq{eq:Apol_lorentz} does not provide an accurate characterization of these absorption spectra. In relation to these methods, we equate the frequency of the polariton peak maxima with $\omega_{1,2}$ and the respective half width at half maximum (HWHM) with $\Gamma_{1,2}$. 

\begin{figure}
\includegraphics[width=\columnwidth]{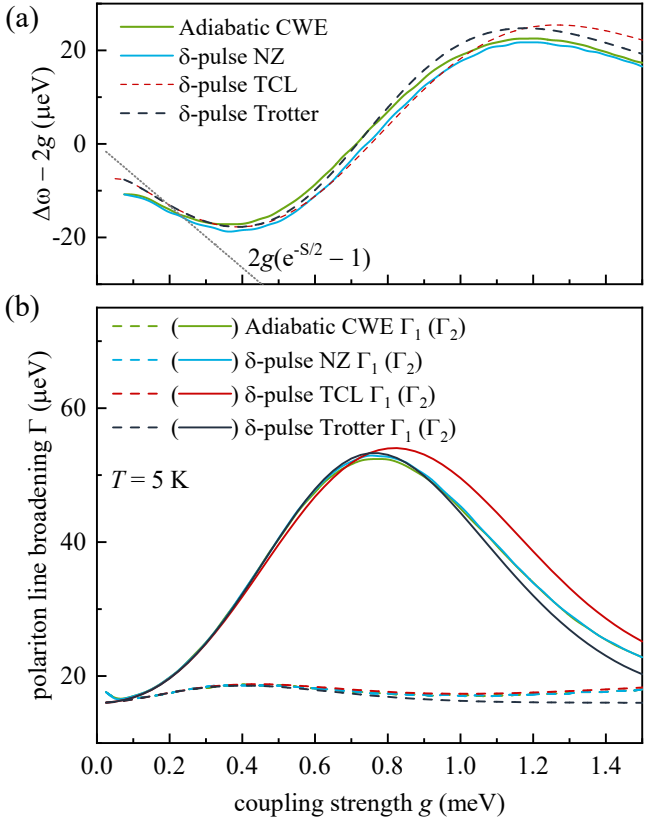}
\caption{(a) Deviation of the polariton Rabi splitting $\Delta \omega = \omega_2 - \omega_1$ from the nominal Rabi splitting $2g$ as a function of the exciton-cavity coupling strength $g$, calculated according to the Adiabatic CWE polaron master equation (solid green), NZ equation (solid blue), TCL equation (dashed red) and TD (dashed black). The deviation of the phonon renormalized Rabi splitting from the nominal Rabi splitting $2g(e^{-S/2}-1)$ is also shown (dotted black). System parameters include $\omega_C=\omega_X + \Omega_p$, $\gamma_X = 2\,\mu$eV, $\gamma_C = 30\,\mu$eV and $T=5$ K. (b) Linewidths $\Gamma_{1,2}$ of the lower (solid lines) and upper (dashed lines) polariton states as functions of the  exciton-cavity coupling strength $g$, calculated according to the four above-described approaches.\label{fig:Gamma_vs_gT5}}
\end{figure}

\begin{figure}
\includegraphics[width=\columnwidth]{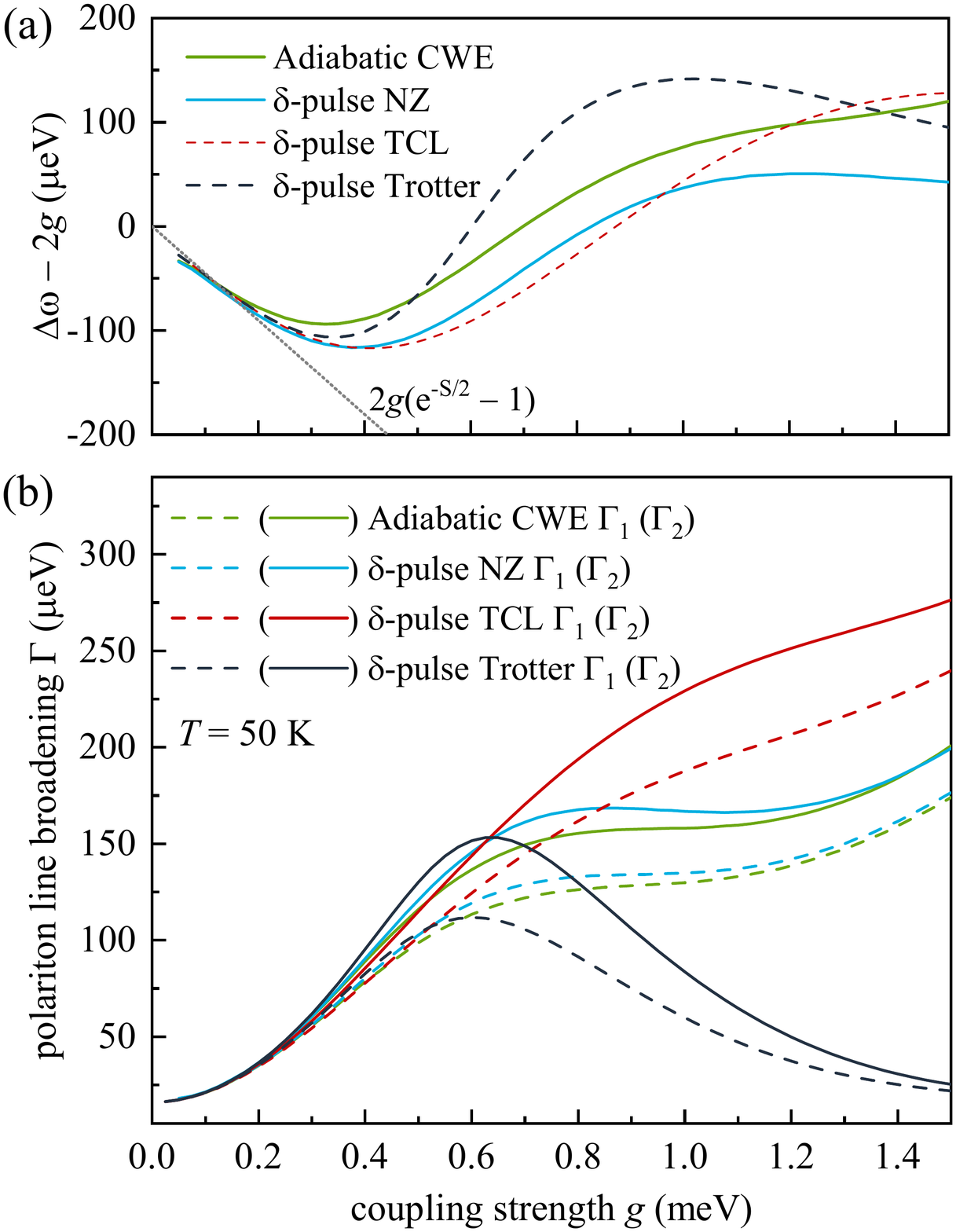}
\caption{As \Fig{fig:Gamma_vs_gT5} but for $T=50$ K. \label{fig:Gamma_vs_gT50}}
\end{figure}

The polariton Rabi splitting $\Delta \omega = \omega_2 - \omega_1$ and linewidths $\Gamma_{1,2}$ 
as calculated according to the three polaron master equation approaches and the TD method, are shown as functions of coupling strength $g$ in \Fig{fig:Gamma_vs_gT5} for $T=5$ K and \Fig{fig:Gamma_vs_gT50} for $T=50$ K. Clearly,  the accuracy of all three polaron master equation techniques deteriorates with increasing exciton-cavity coupling strength $g$. This is a consequence of the approximation common for all of these techniques, which neglects any terms in the master equation higher than $g^2$.  This failure of the master equation approaches is getting significantly more pronounced at higher temperatures, owing to the enhanced influence of the phonon environment.

The regimes of validity of all polaron master equation approaches are limited by the magnitude of the perturbative term $\mathcal{H}_{\rm int} / \mathcal{H}_{\rm int}^{\delta}$ [\Eq{eq:Hprime_int}/(\ref{eq:Hprime_int_delta})]. Following similar argumentation to that outlined in Ref.\,\cite{McCutcheonNJP10}, we estimate the condition for validity of the second order Born approximation to be
\begin{equation}
\left(\frac{g}{\omega_0}\right)^2 (1- \langle B \rangle^4) \ll 1\,,\label{eq:pert_cond}
\end{equation}
where $g$ is the exciton-cavity coupling strength, $\omega_0$ is the cut-off frequency of the phonon spectral density $J(\omega)$ [see \Eq{eq:Jomega}] and $\langle B \rangle$ is the expectation value of the phonon bath displacement operators [see \Eq{eq:Bavg}]. The condition defined in \Eq{eq:pert_cond} may be alternatively expressed in terms of the polariton and polaron timescales [Eqs. \ref{eq:tauJC} and \ref{eq:tauIB}, respectively], with $g/\omega_0 \sim \tau_{\rm IB}/\tau_{\rm JC}$.

For the selected parameter set, $\langle B \rangle$ is equal to $0.97$ at $T=5$ K or $0.77$ at $T=50$ K. As a very crude estimate, we therefore expect the maximum coupling strength $g$ at which the polaron master equations remain valid to be approximately five times greater at $T=5$ K compared to $T=50$ K. This estimate is fully compatible with the behavior of the polariton line parameters illustrated in Figs. \ref{fig:Gamma_vs_gT5} and \ref{fig:Gamma_vs_gT50}.


\section{Conclusion}
\label{sec:conclusion}

The broadband is practically absent in the pulsed excitation regime, which is a significant drawback of the mater equation approach. The Markov approximation, which introduces a difference between NZ and TCL approaches, does not meaningfully change the results, as the main memory effect contributes via the broadband which is ruined by the density matrix factorization -- the key assumption of the master equation approach. Furthermore, the master equation approach fails with increasing QD-cavity coupling strength $g$ and temperature, as it can be correct only up to second order in $g$, so overall it does not demonstrate significant advantages relative to the fully analytic approach.


The TD shares many parameters with the polaron master equation techniques. Indeed, the only additional computational steps required for implementation of the TD relative to the polaron master equation approaches relate to calculation of matrix products. The TD method therefore benefits from the computational simplicity of these master equation approaches but provides the universal validity (for example very strong coupling strengths and high temperatures) of more computationally complex path integral techniques.

\begin{acknowledgments}
The authors acknowledge support by the EPSRC under the DTA
scheme and grant EP/M020479/1. The authors thank W. Langbein and  I. Wilson-Rae for discussions.
\end{acknowledgments}

\appendix

\section{Link between the linear polarization and absorption}
\label{sec:pulsed_abs}
In this Appendix, we provide a detailed derivation of \Eq{eq:Absorp_pulsed}.
We take, as our starting point, the standard definition of optical polarization given in \Eq{eq:genericP}. After the initial pulsed excitation is applied, the density matrix $\rho(t)$ evolves according to the standard Schr\"odinger-representation time evolution,
\begin{equation}
\rho(t>0) = e^{-iHt} \rho(0_+) e^{iHt}\,,\label{eq:rhotgt0}
\end{equation}
where $\rho(0+)$ is given by \Eq{eq:rho0+}. If we consider only linear polarization we may neglect all terms higher than first order in $\Omega_e$. Thus, substituting for $\rho(t)$ in \Eq{eq:genericP}, the linear polarization $P_L(t)$ has the form,
\begin{equation}
P_L(t) = \tr\{e^{-iHt}\, c_e^{\dagger} \, \rho(-\infty) \,e^{iHt} \,c_o\},\label{eq:linearpol}
\end{equation}
where we have dropped the factor of $-i$ and normalized to excitation strength $\Omega_e$.

To find the absorption in the regime of CWE, we apply the basic principle of conservation of probability. In terms of formal scattering theory, the absorption $A(\omega)$ is the rate for transitions into all states other than the initial state. We employ Fermi's golden rule, which states that for a harmonic perturbation $\mathcal{V}(t) = \Omega_e c_e e^{-i\omega t}$ the probability of a transition from initial state $i$ to final state $f$ occurring per unit time, $R_{i \to f}$, is given by,
\begin{equation}
R_{i \to f}(\omega) = 2 \pi |\bra{i} \Omega_e  c_e \ket{f}|^2 \delta(E_f - E_i - \omega),
\end{equation}
where $E_{i(f)}$ is the energy of the initial (final) state $\ket{i}$ ($\ket{f}$).

The absorption is given by the transition rate between all initial and final states, with each contribution weighted by the probability that the system is found in the associated initial state $w_i$
\begin{align}
A(\omega) &= \sum_{i, f\neq i} w_i \, \, \Omega_e ^2 \, R_{i \to f}(\omega)\\
&= 2\pi \sum_{i,f \neq i} w_i \, \Omega_e^2 \, |\bra{i} c_e \ket{f}|^2 \delta(E_f - E_i - \omega).\label{eq:Fermiabs}
\end{align}

Noting that a general delta-function in the frequency domain $\delta(\omega)$ may be written as,
\begin{equation}
\delta(\omega) = \frac{1}{2\pi} \int_{-\infty}^{\infty} dt \, e^{-i\omega t},
\end{equation}
we recast \Eq{eq:Fermiabs} as,
\begin{align}
A(\omega) &= \Omega_e^2 \int_{-\infty}^{\infty} dt \, \sum_{i,f \neq i} w_i |\bra{i} c_e \ket{f}|^2 e^{-i(E_f - E_i)t}e^{i\omega t}\nonumber\\
&\hspace{-0.5cm} = \Omega_e^2\int_{-\infty}^{\infty} dt \, \sum_{i,f \neq i} w_i \bra{i} e^{i H t} \, c_e \, e^{-i H t} \ket{f} \bra{f} c_e^{\dagger} \ket{i} e^{i\omega t}\nonumber\\
&\hspace{-0.5cm} = \Omega_e^2 \int_{-\infty}^{\infty} dt \, \sum_{i} w_i \bra{i} e^{i H t} \, c_e \, e^{-i H t} \, c_e^{\dagger} \ket{i} e^{i\omega t}\,,\label{eq:abs_intermediate}
\end{align}
where in the last equality we have used the fact that the perturbation $c_e^{\dagger}$ does not contain any diagonal elements and thus terms with $f=i$ provide no contribution.  Recalling that $w_i$ is the probability of finding the system in initial state $i$ before the pulse is applied, $w_i = \bra{i} \rho(-\infty) \ket{i}$, and noting that $\sum_i \bra{i} \ldots \ket{i}$ describes the trace operation, \Eq{eq:abs_intermediate} becomes
\begin{align}
A(\omega) &= \Omega_e^2 \int_{-\infty}^{\infty} dt \, e^{i\omega t} \tr\left\{e^{-iHt} \, c_e^{\dagger} \, \rho(-\infty) \, e^{iHt} \, c_e \right\},
\end{align}
where we have used the cyclic property of the trace operation. Extending this formalism to a perturbation of the form $\mathcal{V}(t) = \Omega_e (c_e e^{-i\omega t} + c_e^{\dagger} e^{i\omega t})$, the absorption becomes,
\begin{align}
A(\omega) &= \Omega_e^2 \int_{-\infty}^{\infty} dt \, e^{i\omega t} \tr\left\{e^{-iHt} \, c_e^{\dagger} \, \rho(-\infty) \, e^{iHt} \, c_e \right\} + \text{c.c}\,, \label{eq:A_full}
\end{align}
where ${\rm c.c.}$ denotes the complex conjugate of the preceding term. The trace within \Eq{eq:A_full} is in fact the linear polarization given in \Eq{eq:linearpol} with $c_o = c_e$. Dropping the unimportant factor of $2 \Omega_e^2$, we arrive at \Eq{eq:Absorp_pulsed}.

\section{Derivation of second order Born master equation [\Eq{eq:Lindpolaron}]}
\label{app:2ndborn}
We take, as our starting point, the standard Lindblad master equation [\Eq{eq:Lind_master}]. It is helpful to transform from the Schr\"odinger representation to the interaction representation, the latter being defined such that
\begin{equation}
\tilde{{O}}(t) = e^{i\mathcal{H}_0 t} \, {O} \, e^{-i\mathcal{H}_0 t}
\end{equation}
where ${O}$ is a generic operator in the Schr\"odinger representation, $\tilde{{O}}(t)$ is its counterpart in the interaction representation, and $\mathcal{H}_0 = \mathcal{H}_{\rm sys}^{(\delta)} + \mathcal{H}_{\rm ph}$ [with $\mathcal{H}_{\rm sys}^{(\delta)}$ given by Eq. (\ref{eq:Hprime_sys}) or (\ref{eq:Hprime_sys_delta}), according to the method under consideration, and $\mathcal{H}_{\rm ph}$ given by \Eq{eq:HIBcomponents}]. When recast in the interaction representation \Eq{eq:Lind_master} becomes,
\begin{equation}
\frac{d \tilde{\rho}(t)}{d t} = - i[\tilde{\mathcal{H}}_{\rm int}^{(\delta)}(t),\tilde{\rho}(t)] + \tilde{\mathcal{D}}(t),\label{vN_diff_form}
\end{equation}
where $\tilde{\mathcal{H}}_{\rm int}^{(\delta)}(t)$ is the interaction representation of \Eq{eq:Hprime_int}/(\ref{eq:Hprime_int_delta}). \Eq{vN_diff_form} has the formal solution
\begin{equation}
\tilde{\rho}(t) = \tilde{\rho}(t_0) + \int_{t_0}^t d\tau \left(-i[\tilde{\mathcal{H}}_{\rm int}^{(\delta)}(\tau),\tilde{\rho}(\tau)] + \tilde{\mathcal{D}}(\tau)\right)\,.\label{vN_int_form}
\end{equation}
Inserting \Eq{vN_int_form} for $\tilde{\rho}(t)$ into \Eq{vN_diff_form} we find that,
\begin{align}
\frac{d \tilde{\rho}(t)}{d t} &= - i\Big[\tilde{\mathcal{H}}_{\rm int}^{(\delta)}(t),\rho(t_0)\Big] -i\left[\tilde{\mathcal{H}}_{\rm int}^{(\delta)}(t),\int_{t_0}^t d\tau \,\,\tilde{\mathcal{D}}(\tau)\right] \nonumber\\
&\hspace{0.5cm}- \left[\tilde{\mathcal{H}}_{\rm int}^{(\delta)}(t),\int_{t_0}^t d\tau \,\, [\tilde{\mathcal{H}}_{\rm int}^{(\delta)}(\tau),\tilde{\rho}(\tau)]\right] + \tilde{\mathcal{D}}(t)\,.\label{eq:rho_tilde_full1}
\end{align}
We apply the {\it weak coupling limit}, which presumes $H_{\rm int}$ to be a small perturbation and therefore enables termination of this iterative procedure at second order [the {\it second-order Born approximation}]. We are interested in the evolution of the polaron-cavity system, and thus require an equation that characterizes the behavior of the reduced density operator $\rho_{S'}(t)$. We therefore take the partial trace of \Eq{eq:rho_tilde_full1} over all bath degrees of freedom,
\begin{align}
\frac{d \tilde{\rho}_{S'}(t)}{d t} &= \tr_B\left\{- i[\tilde{\mathcal{H}}_{\rm int}^{(\delta)}(t),\tilde{\rho}(t_0)]\right\}\nonumber\\
&\hspace{0.5cm}- \tr_B\left\{i\left[\tilde{\mathcal{H}}_{\rm int}^{(\delta)}(t),\int_{t_0}^t d\tau \,\,\tilde{\mathcal{D}}(\tau)\right]\right\}\nonumber\\
 &\hspace{0.5cm}- \int_{t_0}^t d\tau \,\,\tr_B [\tilde{\mathcal{H}}_{\rm int}^{(\delta)}(t),[\tilde{\mathcal{H}}_{\rm int}^{(\delta)}(\tau),\tilde{\rho}(\tau)]] + \tilde{\mathcal{D}}_{S'}(t)\,,\label{eq:vN_intpic}
\end{align}
where $\tilde{\mathcal{D}}_{S'}(t) = \tr_B\{\tilde{\mathcal{D}}(t)\}$ and  $\rho_{S'}(t) = \tr_B \{\rho(t)\}$, with the traces taken over all phonon states in the polaron frame.

Assuming that the bath is sufficiently large to be unaffected by the interaction with the system, we may factorize the polaron frame density matrix $\tilde{\rho}(t)$ at all times,
\begin{equation}
\tilde{\rho}(t) = \tilde{\rho}_{S'}(t) \otimes \tilde{\rho}_{\rm ph}\,,
\end{equation}
where the bath density matrix $\tilde{\rho}_B$ is independent of time. This approximation causes the first and second terms on the RHS of \Eq{eq:vN_intpic} to vanish~\cite{MorreauThes}. Expressing the interaction density matrix $\tilde{\rho}_{S'}(t)$ in terms of the Schr{\"o}dinger density matrix $\rho_{S'}(t)$, we have
\begin{align}
\frac{d \tilde{\rho}_{S'}(t)}{d t} &= \frac{d}{dt}\left(e^{i \mathcal{H}_0 t} \rho_{S'}(t) e^{-i\mathcal{H}_0 t}\right)\nonumber\\
&= e^{i\mathcal{H}_0 t} \frac{d \rho_{S'}(t)}{dt} e^{-i\mathcal{H}_0 t} + i[\mathcal{H}_0,\tilde{\rho}_{S'}(t)]\,,\label{eq:drhotildesdt}
\end{align}
with $\mathcal{H}_0 = \mathcal{H}_{\rm sys}^{(\delta)} + \mathcal{H}_{\rm ph}$, as before. Replacing the LHS of \Eq{eq:vN_intpic} with \Eq{eq:drhotildesdt}, we obtain
\begin{align}
\frac{d \rho_{S'}(t)}{d t} &= -i[\mathcal{H}_{\rm sys}^{(\delta)},\rho_{S'}(t)] + \mathcal{D}_{S'}(t)\nonumber\\
&\hspace{-1cm}- \int_{t_0}^t d\tau \,\,\tr_B [\mathcal{H}_{\rm int}^{(\delta)},e^{-i\mathcal{H}_{0} t}[\tilde{\mathcal{H}}_{\rm int}^{(\delta)}(\tau),\tilde{\rho}_{S'}(\tau)\otimes\rho_B]e^{i\mathcal{H}_{0} t}]\,.
\end{align}
Expressing $\mathcal{H}_{\rm int}^{(\delta)}$ explicitly according to \Eq{eq:Hprime_int}/(\ref{eq:Hprime_int_delta}) and making the change of variables $\tau \to t' = t - \tau$, we arrive at \Eq{eq:Lindpolaron}.


\section{Absorption under adiabatic CWE}
\label{app:CWE}

Here, we provide some intermediate steps of the derivation of the exciton and cavity absorption in the CWE regime.
From Eqs. (\ref{eq:abs_inf}) and (\ref{eq:rhodot_CWE_red}), we obtain
\begin{equation}
A(\omega) = {\cal Q}_{00} + \begin{pmatrix} {\cal Q}_{0X} & {\cal Q}_{0C}\end{pmatrix} \begin{pmatrix} \rho_{X0}(\infty) \\ \rho_{C0}(\infty) \end{pmatrix} + \text{H.c.}\,,
\end{equation}
with $\rho_{j0}(\infty)$ determined by the relation,
\begin{equation}
\begin{pmatrix} {\cal Q}_{X0} \\ {\cal Q}_{C0} \end{pmatrix} = \underbrace{\begin{pmatrix} {\cal Q}_{XX} & {\cal Q}_{XC} \\ {\cal Q}_{CX} & {\cal Q}_{CC} \end{pmatrix}}_{\hat{{\cal Q}}_R} \begin{pmatrix} \rho_{X0}(\infty) \\ \rho_{C0}(\infty) \end{pmatrix}\,,\label{eq:QR_def}
\end{equation}
and elements ${\cal Q}_{jk}$ defined by \Eq{eq:Qmatrix}.

It is instructive, at this stage, to separate excitonic and cavity feeding channels. Substituting for ${\cal Q}_{jk}$ from \Eq{eq:Qmatrix}, we find 
\begin{align}
A(\omega) &= \Omega_X^2 \left(\mathcal{W}_{XX} + \hat{f}_{X}^T \, \hat{Q}_R^{-1} \hat{f}_{X} + {\rm c.c.}\right)\nonumber\\
&\hspace{-0.5cm} + \Omega_C^2 \left( \hat{f}_{C}^T \, \hat{Q}_R^{-1} \hat{f}_{C} + {\rm c.c.}\right) + {\rm cross \, terms}\,, \label{eq:abs_full_CW}
\end{align}
where $\hat{Q}_R$ is defined in \Eq{eq:QR_def} and $\hat{f}_{X,C}$ are vectors associated with the exciton and cavity excitation modes,
\begin{equation}
\hat{f}_{X} = \begin{pmatrix} \langle B \rangle - ig\mathcal{W}_{CX} \\ -ig \mathcal{W}_{XX}\end{pmatrix}\,,\hspace{0.8cm}
\hat{f}_{C} = \begin{pmatrix}0\\ 1 \end{pmatrix}\,.
\end{equation}
The terms prefixed by factor $\Omega_{X}^2$ ($\Omega_{C}^2$) describe absorption under CWE in the exciton (cavity) mode; the cross terms (prefixed by $\Omega_X \Omega_C$) have not been included explicitly since these terms cannot be physically interpreted as absorption. 

To find absorption associated with excitation in the exciton mode $A_X(\omega)$, we set $\Omega_C$ to zero within \Eq{eq:abs_full_CW}. Dropping the scaling factor of $\Omega_X^2$, we arrive at \Eq{eq:abs_X_CW}. An equivalent procedure is applied to find absorption associated with excitation in the cavity mode $A_C(\omega)$.

\section{Special case: zero detuning} 
\label{app:zerodet}


In this section we compare the above-described approaches in the particular case of zero effective detuning $\bar{\omega}_X = \omega_C$ and equal long-time ZPL dephasing and radiative decay rates $\gamma_X = \gamma_C$. In this special case, the matrix $\bar{H}_{\rm JC}$ given by \Eq{eq:QR} is diagonalized as follows:
\begin{equation}
\bar{H}_{\rm JC} = \frac{1}{2} \begin{pmatrix} 1 & -1 \\ 1 & 1\end{pmatrix} 
\begin{pmatrix} \omega_C - \bar{g} &0 \\ 0 & \omega_C + \bar{g} \end{pmatrix}\begin{pmatrix} 1 & 1 \\ -1 & 1\end{pmatrix} \,.
\label{eq:Hdiag}
\end{equation}

\subsection{Adiabatic CWE solved by polaron master equation [\Sec{sec:WR}]}
For the zero detuning, the matrix elements $W_{jk}(t)$ given by \Eq{eq:Wjk_t} may be expressed as
\begin{equation}
W_{jk}(t) = \begin{dcases}
e^{-i\omega_C t} \cos(\bar{g}t) \, G_+(t) & \text{for}\, j=k\,,\\
-i\, e^{-i\omega_C t} \sin(\bar{g}t) \, G_-(t) & \text{for}\, j\neq k\,,\\
\end{dcases}
\end{equation}
with $G_{\pm}(t)$ given by \Eq{eq:Gpm}. Accordingly, elements $\mathcal{W}_{jk}(\omega) = \int_0^{\infty} dt \, e^{i\omega t} \, W_{jk}(t)$ have the form
\begin{equation}
\mathcal{W}_{jk}(\omega) = \begin{dcases}
\sum_{\eta = \pm 1} \mathcal{G}_+(\omega - \omega_C + \eta \bar{g}) & \text{for}\, j=k\,,\\
\sum_{\eta = \pm 1} -\eta \, \mathcal{G}_-(\omega - \omega_C + \eta \bar{g})
& \text{for}\, j\neq k\,,\\
\end{dcases}
\end{equation}
where $\mathcal{G}_{\pm}(\omega) = \int_0^{\infty} dt \, G_{\pm}(t) e^{i\omega t}$ is the Fourier-Laplace transform of $G_{\pm}(t)$.

Noting that in the zero detuning case $\mathcal{W}_{XX} = \mathcal{W}_{CC}$ and $\mathcal{W}_{CX} = \mathcal{W}_{XC}$, we now define the following:
\begin{equation}
\mathcal{W}_{\pm}(\omega) = \mathcal{W}_{XX}(\omega) \pm \mathcal{W}_{CX}(\omega)\,.
\end{equation}

From \Eq{eq:abs_X_CW}, absorption in the excitonic (cavity) mode $A_X(\omega)$ ($A_C(\omega)$) in the adiabatic CWE regime is therefore given by
\begin{align}
A_X(\omega) &= \Re \left\{\mathcal{W}_{XX} + \frac{\langle B \rangle^2}{2}\left(\frac{1}{\lambda_+} + \frac{1}{\lambda_-}\right)\right.\nonumber\\
&\hspace{0.5cm} \left.+ i\bar{g}\left(\frac{\mathcal{W}_-}{\lambda_+} - \frac{\mathcal{W}_+}{\lambda_-}\right)
- \frac{g^2}{2}\left(\frac{\mathcal{W}_-^2}{\lambda_+} + \frac{\mathcal{W}_+^2}{\lambda_-}\right) \right\}\,,\label{eq:AX_CWE_zerodetuning}\\
A_C(\omega) &= \Re \left\{\frac{1}{2}\left(\frac{1}{\lambda_+} + \frac{1}{\lambda_-}\right)\right\}\,,\label{eq:AC_CWE_zerodetuning}
\end{align}
with 
\begin{equation}
\lambda_{\pm}(\omega) = -i(\omega - \omega_C  \pm \bar{g}) + g^2\mathcal{W}_{\pm}(\omega)\,,\label{eq:lambdapm}
\end{equation}
the eigenvalues of matrix $\hat{\cal Q}_R(\omega)$ defined in \Eq{eq:calQR}. Note that, owing to 
$\bar{\omega}_X = \omega_C$ and $\gamma_X = \gamma_C$, matrices $\hat{\cal Q}_R(\omega)$  and $\bar{H}_{\rm JC}$ are diagonalized by the same transformation \Eq{eq:Hdiag}.


\subsection{Pulsed excitation solved by NZ polaron master equation}
Taking the expressions for $\mathcal{W}_{jk}$ above, absorption in the exciton mode following a pulsed excitation is given by the NZ approach as,
\begin{equation}
A_X(\omega) = \Re \left\{\frac{\langle B \rangle^2}{2}\left(\frac{1}{\lambda_+} + \frac{1}{\lambda_-}\right)\right\}\,,
\end{equation}
where $\lambda_{\pm}$ are defined in \Eq{eq:lambdapm}. Absorption in the cavity mode $A_C(\omega)$ is identical to that found from the adiabatic CWE regime [\Eq{eq:AC_CWE_zerodetuning}].

\subsection{Pulsed excitation solved by TCL polaron master equation}
\begin{align}
M_+(t) &= \begin{pmatrix} \sin^2(\bar{g}t) & -i\sin(\bar{g}t)\cos(\bar{g}t)\\-i\sin(\bar{g}t)\cos(\bar{g}t) & \sin^2(\bar{g}t) \end{pmatrix}\,,\\
M_-(t) &= \begin{pmatrix} \cos^2(\bar{g}t) & i\sin(\bar{g}t)\cos(\bar{g}t)\\ i\sin(\bar{g}t)\cos(\bar{g}t) & \cos^2(\bar{g}t)\end{pmatrix}\,.
\end{align}

\begin{align}
\hat{Q}_{\rm TCL}(t) &= i\bar{H}_{\rm JC} + \int_{t_0}^t dt' \, g^2 G_g(t') \mathbb{1}\nonumber\\
& + g^2 G_u(t') \begin{pmatrix} \cos(2\bar{g}t') & i\sin(2\bar{g}t') \\ i\sin(2\bar{g}t') & \cos(2\bar{g}t') \end{pmatrix}\,,
\end{align}
with $G_{g,u}(t)$ defined in Eqs. (\ref{eq:Gg_t}) and (\ref{eq:Gu_t}).


%
%
%
%
%
%
%

\section{IB model cumulant}
\label{App:Cumulant}


The IB model cumulant $K(t) $ can be conveniently written in terms of the standard phonon propagator $D_q$,
\be
K(t)= - \frac{i}{2} \int_0^t d\tau_1 \int_0^t d\tau_2 \sum_q |\lambda_q|^2 D_q(\tau_1 - \tau_2)\,,
\label{Cum}
\ee
where
\begin{align}
i D_q(t) &= \langle \mathcal{T} [b_q(t)+b^\dagger_{-q}(t)]^\dagger [b_q(0)+b^\dagger_{-q}(0)]\rangle\nonumber\\
&=N_q e^{i\omega_q |t|} + (N_q + 1) e^{-i\omega_q |t|},
\label{Ddef}
\end{align}
and $N_q$ is the Bose distribution function,
\begin{equation}
N_q = \frac{1}{e^{\beta \omega_q} - 1}.\label{eq:BoseN}
\end{equation}
Performing the integration in \Eq{Cum}, we obtain
\begin{align}
K(t) &= \sum_q |\lambda_q|^2 \nonumber\\
&\times \left(\frac{N_q}{\omega_q^2}\left[e^{i\omega_q t} - 1\right] + \frac{N_q + 1}{\omega_q^2}\left[e^{-i\omega_q t} -1\right] + \frac{it}{\omega_q}\right)\,.
\end{align}
Converting the summation over $q$ to an integration $\sum_q \rightarrow \frac{\Vol}{(2\pi)^3 v_s^3} \int d^3 \omega$, where $\Vol$ is the sample volume, we then have
\begin{align}
K(t) &= \frac{4 \pi \Vol}{(2\pi)^3 v_s^3} \int_0^{\infty} d\omega \, \omega^2 |\lambda_q|^2 \nonumber\\
&\times \left(\frac{N_q}{\omega^2}\left[e^{i\omega t} - 1\right] + \frac{N_q + 1}{\omega^2}\left[e^{-i\omega t} -1\right] + \frac{it}{\omega}\right)\,.
\label{Kt}
\end{align}
Noting that $|\lambda_q|^2$ may be expressed in terms of the spectral density function $J(\omega) = \sum_{q}|\lambda_q|^2 \delta(\omega - \omega_q)$, we re-write \Eq{Kt} as
\begin{align}
K(t) &= \int_0^{\infty} d\omega \, J(\omega) \nonumber\\
&\times \left(\frac{N_q}{\omega^2}\left[e^{i\omega t} - 1\right] + \frac{N_q + 1}{\omega^2}\left[e^{-i\omega t} -1\right] + \frac{it}{\omega}\right) \\
&= - i\Omega_p t - S + \phi(t).\label{eq:fullK}
\end{align}
In the last equality, the IB cumulant is expressed in terms of the polaron shift $\Omega_p$, the Huang-Rhys factor $S$, and the phonon propagator $\phi(t)$, which are given by Eqs.\,(\ref{eq:omegaP_me}), (\ref{eq:Huang_Rhys}), and (\ref{eq:phi_t}), respectively.


\bibliography{comparison_references_EMmod.bib}
\end{document}